\documentclass[aps,prx,superscriptaddress,twocolumn,eprint]{revtex4-2}
\usepackage{amsmath}
\usepackage{amsthm}
\usepackage{mathtools}
\usepackage{hyperref}
\usepackage{amssymb}
\usepackage{physics}
\usepackage{graphicx}
\graphicspath{ {figures/} }
\usepackage[caption=false]{subfig}
\usepackage{bm}
\usepackage[dvipsnames]{xcolor}
\usepackage{float}

\usepackage{setspace} 

\usepackage[section]{placeins} 

\usepackage{comment}

\usepackage{empheq}

\usepackage{tikz}
\usetikzlibrary{positioning, calc}

\newcommand{\p}[2]{\ensuremath{\frac{\partial #1}{\partial #2}}} 

\newcommand{\beq}{\begin{equation}}
\newcommand{\eeq}{\end{equation}}

\begin{document}

\title{Periodic orbits, pair nucleation, and unbinding of active nematic defects on cones}
\author{Farzan Vafa}
\affiliation{Center of Mathematical Sciences and Applications, Harvard University, Cambridge, MA 02138, USA}
\author{David R. Nelson}
\affiliation{Department of Physics, Harvard University, Cambridge, MA 02138, USA}
\author{Amin Doostmohammadi}
\affiliation{Niels Bohr Institute, University of Copenhagen, Blegdamsvej 17, Copenhagen 2100, Denmark}
\date{\today}

\begin{abstract}
    
    Geometric confinement and topological constraints present promising means of controlling active materials. By combining analytical arguments derived from the Born-Oppenheimer approximation with numerical simulations, we investigate the simultaneous impact of confinement together with curvature singularity by characterizing the dynamics of an active nematic on a cone. Here, the Born-Oppenheimer approximation means that textures can follow defect positions rapidly on the time scales of interest. Upon imposing strong anchoring boundary conditions at the base of a cone, we find a a rich phase diagram of multi-defect dynamics including exotic periodic orbits of one or two $+1/2$ flank defects, depending on activity and non-quantized geometric charge at the cone apex. By characterizing the transitions between these ordered dynamical states, we can understand (i) defect unbinding, (ii) defect absorption and (iii) defect pair nucleation at the apex. Numerical simulations confirm theoretical predictions of not only the nature of the circular orbits but also defect unbinding from the apex.
        
\end{abstract}

\maketitle


\section{Introduction}

Nematic (apolar) order is ubiquitous in biological systems, ranging from subcellular filaments~\cite{sanchez2012spontaneous,keber2014topology,zhang2018interplay}, to bacterial biofilms~\cite{dell2018growing,you2018geometry,copenhagen2021topological}, and cell monolayers~\cite{duclos2018spontaneous,blanch2018turbulent}. In two-dimensional nematics, which emerge when head-tail symmetry arsies upon coarse-graining, the lowest energy defects are $\pm 1/2$ disclinations~\cite{gennes1993the}. Notably, topological defects and their dynamics are often used to characterize \emph{active} nematics, which consist of elongated units that consume energy to generate motion~\cite{needleman2017active,doostmohammadi2018active,shankar2022topological}. Moreover, these defects can mediate various biological functions, including cell extrusion and apoptosis in mammalian epithelia~\cite{saw2017topological}, neural mound formation~\cite{kawaguchi2017topological}, bacterial competition~\cite{meacock2021bacteria}, and limb origination in simple animals such as regenerating \emph{Hydra}~\cite{maroudas2021topological}. For recent reviews on the significance of topological defects in biological systems, see Refs.~\cite{doostmohammadi2021physics, shankar2022topological, bowick2022symmetry}.

The ability to control active materials, with their intricate interplay of topological defects, orientational order, flows, and geometry, would provide exciting new possibilities to transport matter, energy, and information far from equilibrium~\cite{shankar2022spatiotemporal,needleman2017active,zhang2021autonomous}. The  control method we consider in this paper is confinement, which has been intensely studied in flat geometries such as channels~\cite{shendruk2017dancing,duclos2018spontaneous,hardouin2019reconfigurable,opathalage2019self,chandragiri2019active,varghese2020confinement,chandragiri2020flow,li2021formation,wagner2022exact,joshi2023disks}, disks~\cite{duclos2017topological,gao2017analytical,norton2018insensitivity,opathalage2019self,mirantsev2021behavior,hardouin2022active,sciortino2023polarity,joshi2023disks}, and annuli~\cite{wu2017transition,shendruk2017dancing,chen2018dynamics,hardouin2019reconfigurable,hardouin2020active,hardouin2022active,de2023orientational,joshi2023disks}. Curved geometries have attracted less attention, with theoretical and experimental studies focusing on spherical~\cite{lubensky1992orientational,keber2014topology,zhang2016dynamic}, ellipsoidal~\cite{alaimo2017curvature,clairand2023dynamics}, and toroidal~\cite{bowick2004curvature,ellis2018curvature} geometries. For a generic surface, Turner and Vitelli~\cite{vitelli2004anomalous} showed that Gaussian curvature gives rise to an effective topological charge density, which interacts with quantized topological defects.

However, these prototypical surfaces do not typically exhibit both non-trivial Gaussian curvature and a boundary that confines the nematic. The simplest geometry that embodies both properties is the cone, which is not only bounded by the base, but also flat everywhere except at the apex, where a delta function Gaussian curvature singularity resides. In the passive context, recent studies analytically determined the ground state defect configurations for liquid crystals on cones for free~\cite{zhang2022fractional} and tangential~\cite{vafa2022defectAbsorption} boundary conditions imposed at the base. With tangential boundary conditions, cone ground states for nematics can display two, one, or zero $+1/2$ defects on the flanks with increasing cone deficit angle~\cite{vafa2022defectAbsorption}. Here, the cone deficit angle $2\pi\chi$ is given by the angular fraction $\chi$ of a flat disk removed to make a particular cone.
        
\begin{figure}[t]
    \centering
    \includegraphics[width=\columnwidth]{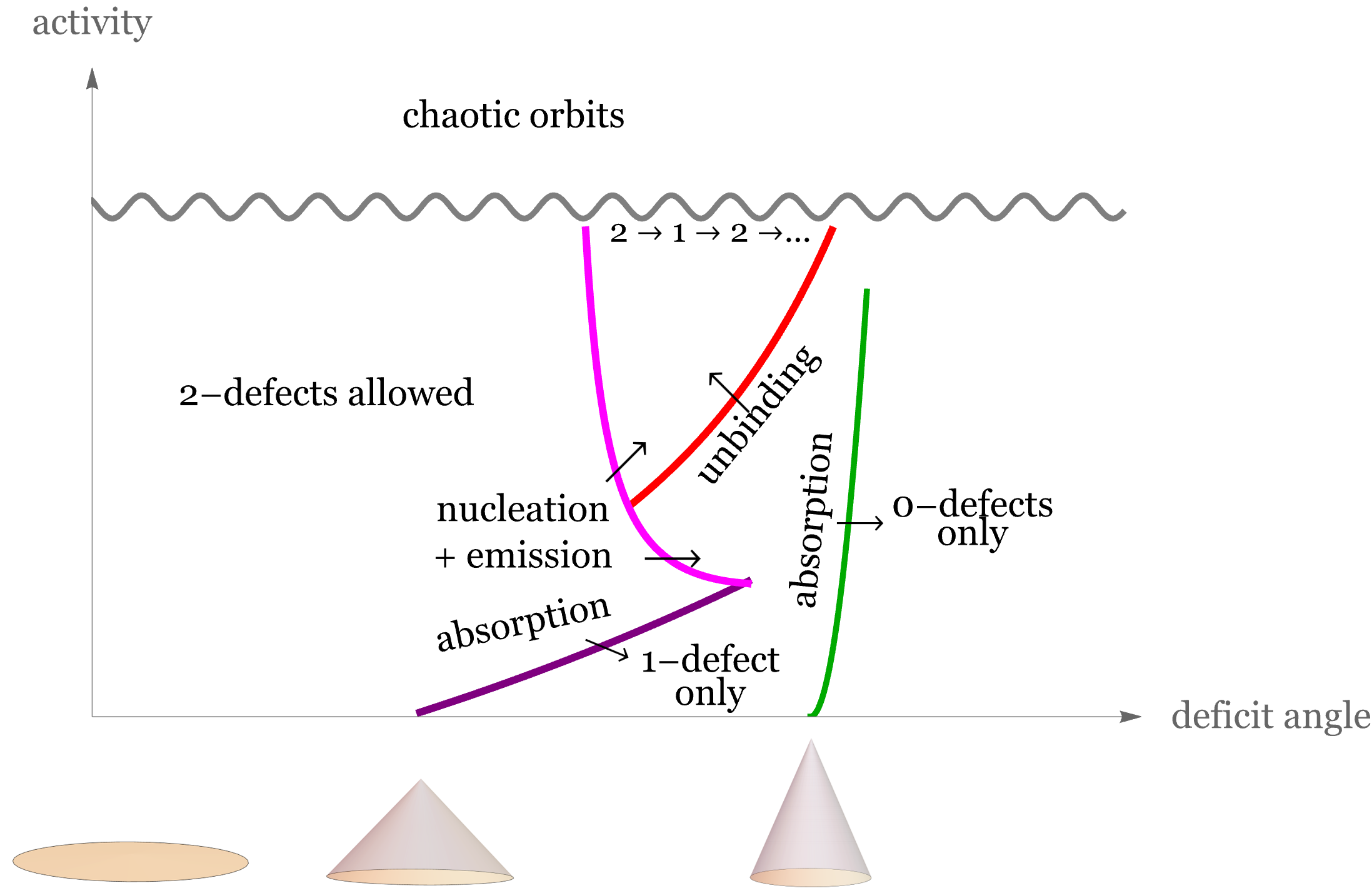}
    \caption{Schematic phase diagram of main results. The colored curves guide the eye. At zero activity, the ground state consists of two flank defects (zero deficit angle, corresponding to a flat disk), one flank defect (deficit angle $2\pi\chi$ between 0 and $2\pi/3$, where the green curve intersects the $x$-axis), and zero flank defects (deficit angle larger than $2\pi/3$)~\cite{vafa2022defectAbsorption}.  A more quantitative summary of our conclusions appears in Fig.~\ref{fig:phaseDiagramData}.}
    \label{fig:phaseDiagram}
\end{figure}

In this paper, we extend our previous work~\cite{vafa2023active} on the dynamics of a single active nematic defect near a curvature singularity to investigate the combined effect of strong anchoring boundary conditions at the cone base and apex curvature on the dynamic organization of active topological defects. By concentrating on regimes where activity allows one or two defects to survive on the cone flanks, we numerically uncover a rich phase diagram of periodic orbits of one or two $+1/2$ flank defects on a cone, with transitions between these states mediated by defect absorption, defect unbinding, or defect pair nucleation at the apex. Zero defect states on the cone flanks are also possible at high deficit angles--see Fig.~\ref{fig:phaseDiagram} for a summary of our main results, as a function of the cone deficit angle and activity. Moreover, we analytically explain many of the features of the phase diagram in Fig.~\ref{fig:phaseDiagram}.

This paper is organized as follows. We begin in Sec.~\ref{sec:formulation} by briefly reviewing a minimal model of a general nematic texture in the one Frank constant approximation on an arbitrarily curved surface. By working deep in the ordered limit and utilizing isothermal coordinates (recently introduced in the context of liquid crystals~\cite{vafa2022active,vafa2022defectAbsorption}, including those with a $p$-fold rotational symmetry), we construct and analyze in Sec.~\ref{sec:stationary} the quasistatic multi-defect solution in the passive setting on a curved surface. Our analysis reveals the presence of metastable states. Isothermal coordinates are particularly powerful for cones, with a delta function of Gaussian curvature, which acts like an unquantized defect charge. Activity enters in Sec.~\ref{sec:dynamics}, where we review the Born-Oppenheimer approximation, which assumes textures rapidly adapt to defect motions, and its consequences, and also present details for the full numerical simulations for nematics on cones. In the following sections, we present our results in order of increasing activity, summarized schematically in Fig.~\ref{fig:phaseDiagram}. We begin by studying the dynamics of stable two $+1/2$ defect orbits and their polarizations in Sec.~\ref{sec:circular}. In Sec.~\ref{sec:oneDefect}, we turn to single defect orbits and study the defect unbinding transition to two defect stable orbits. Finally, in Sec.~\ref{sec:large}, we show that defect pair nucleation followed by emission from the cone apex is another mechanism for the $1\to2$ stable defect orbit transition, and for sufficiently large deficit angle, defect orbits cyclically transition between single and two defect orbits. Throughout the paper, we quantitatively test our theoretical predictions against full numerical simulations of overdamped active nematics on a cone, finding excellent agreement in the regime of low activity and for the location various of defect unbinding transitions. We also point out regimes where the physics is more challenging, such as the more chaotic defect dynamics that arise in our simulations for large activity (see Fig.~\ref{fig:phaseDiagram}). Finally, in Sec.~\ref{sec:discussion}, we summarize our main results and comment on potential experimental realizations of our predicted dynamically ordered defect structures, as well as on the relation to previous work and extensions.

\section{Minimal model in isothermal coordinates}
\label{sec:formulation}

We begin by briefly recalling the framework for describing a nematic texture on a curved surface, following the presentations in Refs.~\cite{vafa2022active,vafa2022defectAbsorption}. This will then be followed by describing the simulation methods. Throughout the paper we discuss theoretical predictions together with the results of full numerical simulations, validating our theoretical assumptions at each step.

\subsection{Metric and nematic tensor}
In two dimensions it is always possible to choose local complex coordinates $z$ and $\bar z$, known as isothermal (or conformal) coordinates, such that the length of the interval squared can be written as~\cite{gauss1822on},
\beq ds^2 = g_{z\bar z} dz d\bar z + g_{\bar z z} d\bar z dz = 2g_{z\bar z}|dz|^2 = e^{\varphi}|dz|^2, \eeq
where $e^{\varphi}$ is the conformal factor that describes position-dependent isotropic stretching, and $g_{z\bar{z}} = e^{\varphi}/2$ is the metric. 

In isothermal coordinates, the nematic order parameter $\mathbf{Q}$ has only two non-zero components: $Q \equiv Q^{zz}$ and $\bar Q \equiv Q^{\bar z \bar z}$, with $Q = (\bar Q)^*$. For ease of notation, let $\nabla \equiv \nabla_z$ and $\bar\nabla \equiv \nabla_{\bar z}$ denote the covariant derivatives with respect to $z$ and $\bar z$, respectively. Covariant derivatives of $\mathbf{Q}$ are simply
\begin{subequations}
    \begin{gather}
        \nabla Q = \partial Q + 2 (\partial \varphi) Q, \qquad \bar \nabla Q = \bar \partial Q, \label{eq:nabla1}\\ 
        \bar\nabla \bar Q = \bar \partial \bar Q + 2 (\bar\partial \varphi) \bar Q, \qquad \nabla \bar Q = \partial \bar Q, \label{eq:nabla2}
    \end{gather}
    \label{eq:nabla}
\end{subequations}
where $\partial \equiv \partial/\partial z$ and $\bar{\partial} \equiv \partial / \partial\bar z$.

\subsection{Conical geometry}
\begin{figure}[t]
    \begin{minipage}[b]{.49\columnwidth}
        \centering
        \subfloat[]{\includegraphics[width=\columnwidth]{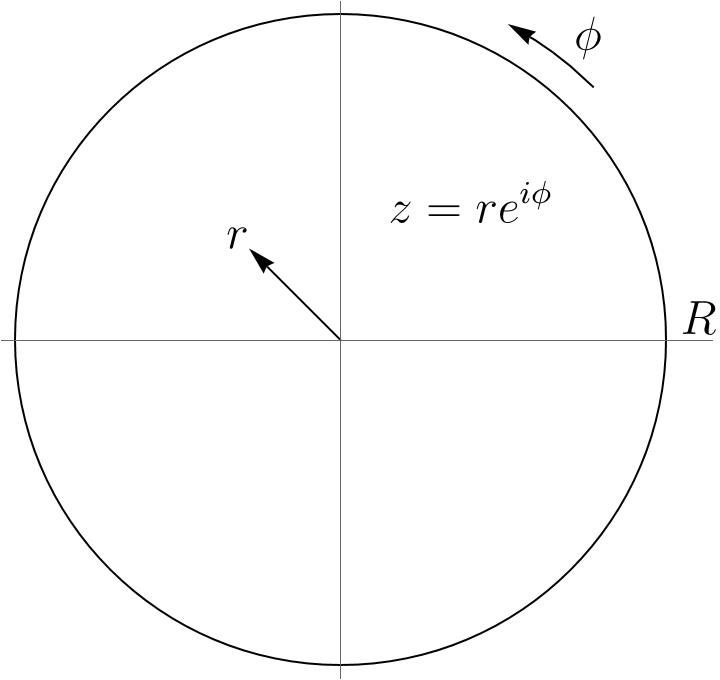}}\\
        \subfloat[]{\includegraphics[width=\columnwidth]{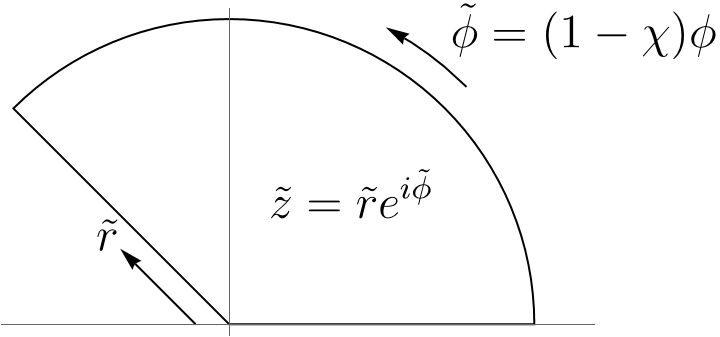}}
    \end{minipage}
    \hfill
    \begin{minipage}[b]{.49\columnwidth}
        \centering
        
        \subfloat[]{\begin{tikzpicture}
                
                \newcommand{\radiusx}{2}
                \newcommand{\radiusy}{.3}
                \newcommand{\height}{5}
                
                \coordinate (a) at (-{\radiusx*sqrt(1-(\radiusy/\height)*(\radiusy/\height))},{\radiusy*(\radiusy/\height)});
                
                \coordinate (b) at ({\radiusx*sqrt(1-(\radiusy/\height)*(\radiusy/\height))},{\radiusy*(\radiusy/\height)});
                
                \draw (a)--(0,\height)--(b);
                
                \draw[dashed] (0,\height)--(0,0);

                \begin{scope}
                    \clip ([xshift=-2mm]a) rectangle ($(b)+(1mm,-2*\radiusy)$);
                    \draw circle (\radiusx{} and \radiusy);
                \end{scope}
                
                \begin{scope}
                    \clip ([xshift=-2mm]a) rectangle ($(b)+(1mm,2*\radiusy)$);
                    \draw[dashed] circle (\radiusx{} and \radiusy);
                \end{scope}
                
                \draw (0,\height-1) arc (-90:-90+atan(\radiusx/\height):1) node[below, pos=.75]{$\beta$}; 
        \end{tikzpicture}}
    \end{minipage}  
    \caption{Schematic of coordinate systems for a cone: (a) Isothermal coordinates $z = re^{i\phi}$. (b) Physical coordinates $\tilde z = \tilde r e^{i\tilde\phi}$, corresponding to an unrolled cone with a missing sector, with angle fraction $\chi$. (c) A diagram of the cone in 3D with cone half angle $\beta$, where $\sin\beta = 1-\chi$.}
    \label{fig:cone}
\end{figure}
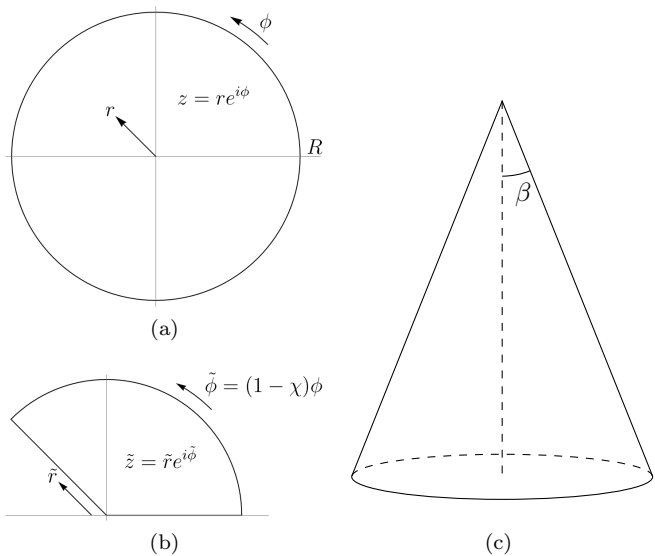

In this paper, we focus on the conical geometry. For a cone with half angle $\beta$, the metric can be obtained from $g_{z\bar{z}}=e^{\varphi}/2$ with
\beq \varphi = -\chi \ln z \bar z, \label{eq:cone} \eeq
where $\chi=1-\sin\beta$. $2\pi\chi$ is known as the deficit angle of the cone, where for example $\chi = 0$ corresponds to a disk (no fraction missing). Another set of useful coordinates are physical coordinates $\tilde z$, which correspond to unrolling a cone to form a planar disk with a missing sector, are related to $z$ via the coordinates 
\beq \tilde z= \frac{z^{1-\chi}}{1-\chi}. \label{eq:ztilde}\eeq
See Ref.~\cite{vafa2022defectAbsorption} for more details and Fig.~\ref{fig:cone} for a schematic of various coordinate systems for a cone.

~\\

\subsection{Free energy}

The Landau-de Gennes free energy $\mathcal F$ on a curved surface can be written as
\beq \mathcal{F} = \int d^2z \sqrt{g}[K |\nabla Q|^2 + K' |\bar \nabla Q|^2 + \epsilon^{-2} (1 - S_0|Q|^2)^2],\label{eq:minimal}\eeq
where explicitly
\begin{align}
    |\nabla Q|^2 &= g_{z \bar z}\nabla Q \bar\nabla \bar Q, \qquad |\bar \nabla Q|^2 = g_{z \bar z}\bar\nabla Q \nabla \bar Q \\
    &\qquad \qquad |Q|^2 = g_{z \bar z}^2 Q \bar Q .  
\end{align} 
Here $K,K' >0$ are Frank elastic type terms~\cite{frank1958liquid,berreman1984tensor}, and in regions of zero Gaussian curvature (such as any point on a cone other than the apex), the two terms are equivalent by integration by parts. The last term governs the isotropic-nematic transition, with $\epsilon$ controlling the microscopic coherence length and $S_0$ sets the equilibrium magnitude of the nematic order. Without loss of generality we set $S_0 = 4$.

Deep in the ordered limit ($\epsilon \ll 1$), the free energy simplifies to~\cite{vafa2022defectAbsorption} 
\beq\mathcal F = J\int d^2z\left|i\partial\alpha + \partial\varphi\right|^2, \label{eq:FSimple}\eeq
where $J = K + K'$ and we have used $Q = e^{-\varphi}e^{i\alpha}$. 

\subsection{Multi-defect solution}

\begin{figure}[t]
    \centering
    \subfloat[]{\includegraphics[width=.495\columnwidth]{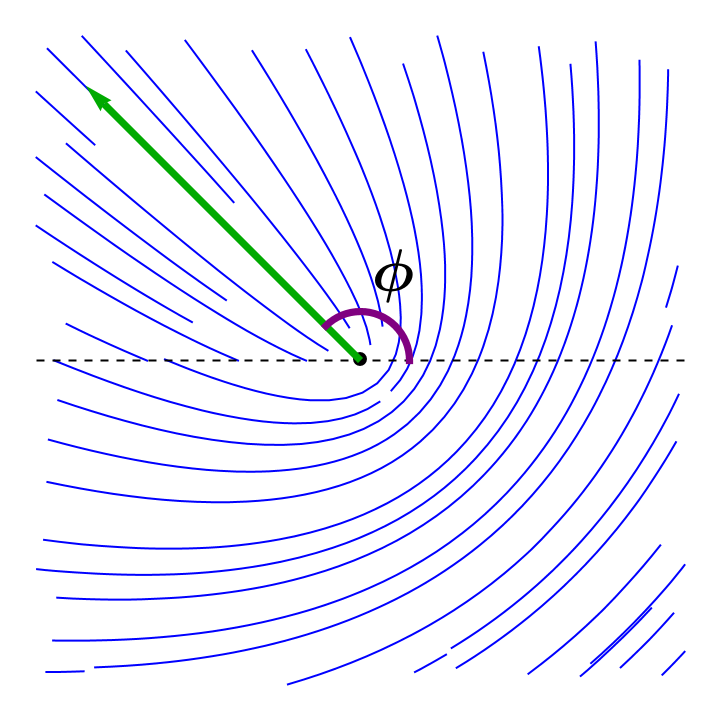}}
    \hfill
    \subfloat[]{\includegraphics[width=.495\columnwidth]{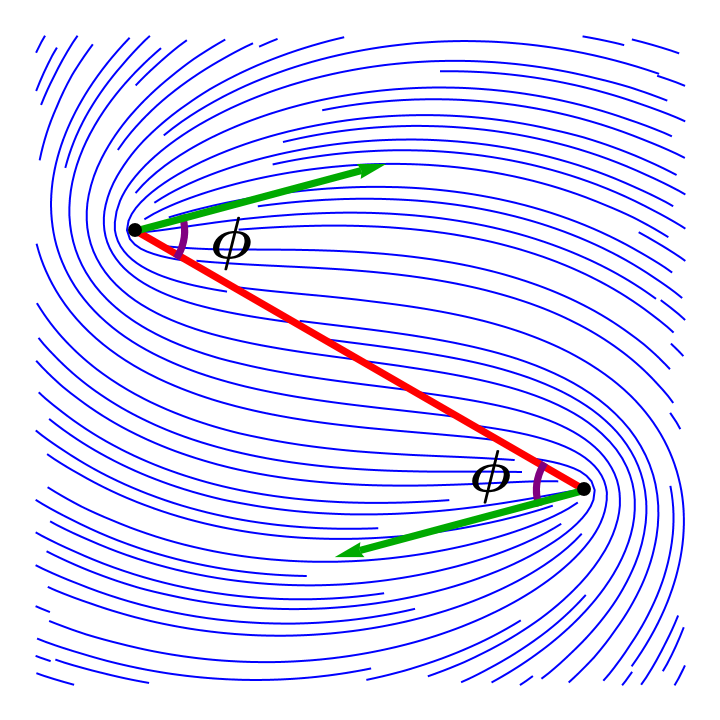}}
    \caption{Sketches of single defect (a) and two defect (b) nematic textures showing the polarization angle $\phi$.}
    \label{fig:polarization}
\end{figure}
The multi-defect ansatz $Q_0$, given by 
\beq Q_0 = e^{i\psi}e^{-\varphi} \prod_j \left(\frac{z - z_j}{|z - z_j|}\right)^{2\sigma_j} \label{eq:Q_0}\eeq
corresponds to minimizing $\mathcal F$ (Eq.~\eqref{eq:FSimple}), with fixed defect positions $z_j$ of charges $\sigma_j$ (which can include appropriate image charges to impose boundary  conditions), and $\psi$ is a constant global phase that defines the overall orientation of the nematic director~\cite{vafa2020multi-defect}. Note that near a defect, e.g. $z \approx z_j$, we can write
\beq Q_0 \approx \left(\frac{z - z_j}{|z - z_j|}\right)^{2\sigma_j} e^{i\phi_j} e^{-\varphi} ,\eeq
where the phase $\phi_j$ defines the polarization of the defect at $z_j$~\cite{vafa2020multi-defect},
\beq e^{i\phi_j} = e^{i\psi}\prod_{j \neq i} \left(\frac{z_i - z_j}{|z_i - z_j|}\right)^{2\sigma_j} . \label{eq:polarization}\eeq
Defects do not have independent polarizations--the polarizations are uniquely determined by all of the defect positions and charges, as well as $\psi$. Note that for a $+1/2$ defect, $\phi_j$ is also the angle of the nematic director.

We now explicitly consider the case of two defects. From the structure of the polarization (Eq.~\eqref{eq:polarization}), it is easy to see that polarizations of neighboring defects are anti-parallel~\cite{vromans2016orientational,vafa2020multi-defect}. For example, for the case of a pair of $+1/2$ defects and $\psi = \pi/2$, then the defect polarizations are perpendicular to the line connecting the two defects and point in opposite directions. See Fig.~\ref{fig:polarization} for sketches of nematic textures for a single $+1/2$ defect and a pair of $+1/2$ defects that show the polarizations. We will return to this observation about the polarization when we consider the dynamics in Sec.~\ref{sec:dynamics}, and then use full numerical simulations in Secs.~\ref{sec:circular} and \ref{sec:large} to determine when this assumption is valid and when it breaks down.

\subsection{Computation of the free energy}

In order to compute the free energy, we must incorporate the boundary conditions. From here onwards, we consider the case of a cone with strong anchoring boundary conditions at the base of radius $R$, by which we mean that the angle of the nematic director relative to the boundary is fixed~\footnote{One can instead consider anti-twist boundary conditions, by which we mean that the nematic director rotates clock-wise around a counter-clockwise loop on the boundary. This means that the net topological charge is $-1$, which means that in the ground state, there will be two anti-podal $-1/2$ defects, just as in the case of tangential boundary conditions for $+1/2$ defects. Unlike $+1/2$ defects, since $-1/2$ defects are not motile, we do not expect the introduction of activity to cause the $-1/2$ defects to orbit (which we checked numerically for small activity independent of the initial condition), although it may be interesting to study this case for large activity.}. To implement this boundary condition, for every defect $j$ of charge $\sigma_j$ at $z_j$, we have a corresponding image defect of charge $\sigma_j$ at $\tilde z_j = R^2/\overline{z_j}$, and using a Gauss' law argument, one can show that the sum of the defect charges is 1~\cite{vafa2022defectAbsorption}. Recalling for a cone that $\varphi = -\chi \ln z \bar z$, computing the free energy (Eq.~\eqref{eq:FSimple}) using the multi-defect solution $Q_0$ (Eq.~\eqref{eq:Q_0}) results in \cite{vafa2022defectAbsorption}
\begin{align}
    \mathcal F &= -2\pi J\left\{\sum_{m<n}\sigma_m\sigma_n\left[\ln \frac{|z_m - z_n|^2}{R^2} + \ln \left|1 - \frac{z_m\overline{z_n}}{R^2}\right|^2 \right] \right. \nonumber\\
    &\qquad \left. + \sum_j\sigma_j^2\ln\left(1 - \frac{|z_j|^2}{R^2}\right) -\chi \sum_j \left(\sigma_j - \frac{\sigma_j^2}{2}\right)\ln \frac{|z_j|^2}{R^2} \right\}\label{eq:FCone}.
\end{align}
The first term (the double sum) represents the elastic interaction between two defects (including image charges). The second term is the self-energy, which would need to be added to any microscopic defect core energy $E_\mathrm{c}$. The final term represents a Coulombic interaction between a topological defect and the geometry~\cite{vitelli2004anomalous}, specialized to the cone. Note that the cone apex develops an effective charge of $-\chi$, and that in interactions with the cone, the effective charge of the defect is modified from $\sigma_j$ to $\sigma_j - \sigma_j^2/2$. Explicitly, for two defects, Eq.~\eqref{eq:FCone} becomes
\begin{align}
    \mathcal F &= -2\pi J\left\{\sigma_1\sigma_2\left[\ln \frac{|z_1 - z_2|^2}{R^2} + \ln \left|1 - \frac{z_1\overline{z_2}}{R^2}\right|^2 \right] \right.\nonumber \\ 
    &\left. {} + \sum_{j=1}^2\sigma_j^2\ln\left(1 - \frac{|z_j|^2}{R^2}\right) -\chi \sum_{j=1}^2 \left(\sigma_j - \frac{\sigma_j^2}{2}\right)\ln \frac{|z_j|^2}{R^2} \right\}\label{eq:F2Defects}.
\end{align}

\subsection{Forces}

Having computed the free energy, we can now compute the $\bar z$-component of the force on defect $i$, i.e. $F_i^{\bar z} = -\p{\mathcal F}{z_i}$, which is given by
\begin{align}
    F_i^{\bar z} &= -2\pi J \left\{ \sigma_i\sum_{j \neq i} \sigma_j\left[ \frac{1}{z_i - z_j} + \frac{1}{z_i - \tilde z_j}\right] + \sigma_i^2 \frac{1}{z_i - \tilde z_i} \right.\nonumber \\
    &\left. \qquad\qquad - \chi \left(\sigma_i - \frac{\sigma_i^2}{2}\right) \frac{1}{z_i}\right\} \nonumber \\ 
    &= -2\pi J \left\{ \sigma_i\sum_{j \neq i} \sigma_j\frac{1}{z_i - z_j} + \sigma_i\sum_j\sigma_j \frac{1}{z_i - \tilde z_j} \right.\nonumber \\
    &\left. \qquad\qquad - \chi \left(\sigma_i - \frac{\sigma_i^2}{2}\right) \frac{1}{z_i}\right\}
\end{align}
All three terms represent Coulombic interactions: the first term, the interaction with all of the other defects; the second term, the interaction with all of the image charges; the third term, the interaction with the apex. Explicitly, for two defects, the force on defect 1 is
\begin{align}
    F_1^{\bar z} &= -2\pi J \left[\sigma_1 \sigma_2\frac{1}{z_1 - z_2} + \sigma_1^2 \frac{1}{z_1 - \tilde z_1} + \sigma_1\sigma_2 \frac{1}{z_1 - \tilde z_2} \right.\nonumber \\
    &\left. \qquad\qquad -\chi \left(\sigma_1 - \frac{\sigma_1^2}{2}\right) \frac{1}{z_1}\right]
    \label{eq:Forces2DefectsCone}
\end{align}

\section{Stationary solutions in passive case on a cone}
\label{sec:stationary}

\subsection{Stability for two defects}

\begin{figure}[t]
    \centering
    \includegraphics[width=\columnwidth]{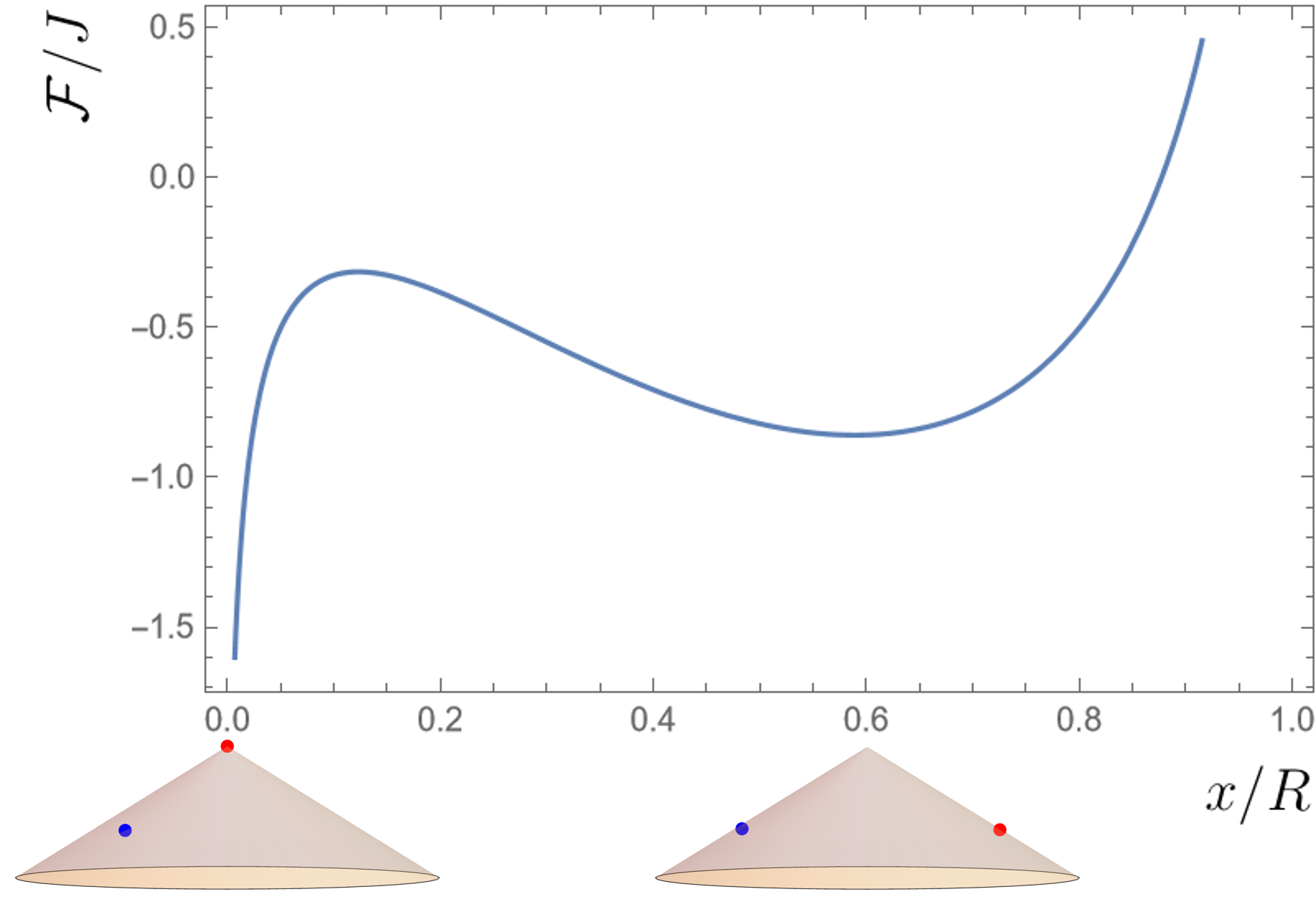}
    \caption{Plot of potential (Eq.~\eqref{eq:F2Defects}) for $\chi = 0.15$ for a pair of $+1/2$ defects, where one defect is at $x_c = 0.59 R$ and the other is at at $x$. The local minimum is at $x = x_c$, but it's not the global minimum as $x \to 0$. Cones below the $x$-axis indicate the solutions at the extrema, where the colored dots indicate the two $+1/2$ defects.}
    \label{fig:2defectPotential}
\end{figure}

We know that for a passive nematic texture on a cone with $0 < \chi < 2/3$, there exists a stable solution where one $+1/2$ defect is at the apex and the other is on the flanks~\cite{vafa2022defectAbsorption}. On the other hand, for the case of nematic texture on a disk ($\chi=0$), a stable solution is two anti-podal $+1/2$ defects~\cite{duclos2017topological,vafa2022defectAbsorption}, in excellent agreement with experiments of fibroblasts on disks~\cite{duclos2017topological}. Thus it is clear that starting from this solution, for sufficiently small $\chi$, by continuity there should exist a classically stable solution of two anti-podal $+1/2$ defects. We now explicitly show this. By computing the eigenvalues of the $4\times 4$ Hessian $H_{ij} = \p{}{x_i}\p{}{x_j} \mathcal F$ at the extremum $x_c = \pm \left(\frac{1-3 \chi }{5-3 \chi }\right)^{1/4}R$ for a pair of $+1/2$ defects, where $\mathcal F$ is given in Eq.~\eqref{eq:FCone}, we find that $x_c$ is a local minimum for $0<\chi < 1-\frac{1}{3} \sqrt{5-\frac{1}{\sqrt{2}}}  = 0.31$. We thus learn that on a cone, there are two classically stable stationary solutions for small enough $\chi$: one solution has one $+1/2$ defect at the apex with another $+1/2$ defect on the flanks, while the second solution has two $+1/2$ defects on the flanks. The global minimum of $\mathcal F$ is obtained for the former. See Fig.~\ref{fig:2defectPotential} for a plot of the potential (Eq.~\eqref{eq:F2Defects}) for a pair of $+1/2$ defects that graphically demonstrates that the two flanks defect configuration is a local but not global minimum.

\section{Dynamical equation}
\label{sec:dynamics}
The minimal model presented thus far describes statics of passive nematic defects on curved geometries. We now introduce dynamics of the defects, explicitly accounting for the effect of active stress generation by the nematogens.

We assume that the motion of the nematic order parameter is controlled by the balance of relaxational dynamics and advection of the tensorial order parameter by flow $v$, according to
\beq \gamma D_t Q = -\frac{1}{\sqrt{g}} g^{z\bar z} g^{z \bar z } \frac{\delta \mathcal F}{\delta \bar Q} \label{eq:QDynamics}\eeq
where
\beq D_t Q = \partial_t Q + (v\nabla + \bar v \bar\nabla) Q - (\nabla v - \bar\nabla \bar v)Q \label{eq:D_tQ}\eeq
is the generalized advective derivative of $Q$, accounting for both regular advection by the flow and reorientation response to flow gradients~\cite{bonn2022fluctuation,vafa2023active} and
\begin{align}
&-\frac{1}{\sqrt{g}} g^{z\bar z} g^{z \bar z } \frac{\delta \mathcal F}{\delta \bar Q} \nonumber \\
&\qquad = g^{z\bar z} \left(K\nabla \bar\nabla + K' \bar \nabla \nabla\right) Q + 2\epsilon^{-2}S_0(1 - S_0|Q|^2)Q\label{eq:num}
\end{align}
is the molecular field.

Assuming that the flow is generated by active stresses and working in the overdamped limit~\cite{doostmohammadi2016stabilization,putzig2016instabilities,srivastava2016negative,oza2016antipolar}, the balance of the active force and the frictional damping leads to
\beq \mu v = \zeta_Q\nabla Q,\eeq
where $\mu$ is the friction coefficient and $\zeta_Q$ is the scalar activity coefficient that characterizes the strength of the active stress. We can thus write
\beq v = \frac{\zeta_Q}{\mu}\nabla Q = \zeta \nabla Q = \zeta \left(\partial\varphi + i\partial\alpha\right)Q,\nonumber\label{eq:viso} \eeq
where $\zeta = \zeta_Q/\mu$.

\subsection{Born-Oppenheimer approximation}

Since the multi-defect ansatz is a stationary solution in the passive setting, then we expect that for small activity, the multi-defect ansatz is still a good solution, provided that the defects are allowed to move, but slowly. Explicitly, we assume
\beq Q_0 = e^{i\psi}e^{-\varphi} \prod_j \left(\frac{z - z_j(t)}{|z - z_j(t)|}\right)^{2\sigma_j} \eeq
where the nematic texture instantaneously readjusts itself in response to the slow motion of the defects. This is known as the Born-Oppenheimer approximation when studying the quantum mechanics of light-weight electrons bonding atoms with much heavier nuclei~\cite{landau2013quantum}.

\subsection{Forces and polarization for a pair of $+1/2$ defects}

We now consider the implications of the Born-Oppenheimer approximation for the prime set-up of two defects on a cone that we focus in this paper. Within the Born-Oppenheimer approximation, the Coulombic forces on a pair of defects (Eq.~\eqref{eq:Forces2DefectsCone} derived in Sec.~\ref{sec:formulation} are still correct, and the motile force on a $+1/2$ defect $j$ is given by~\cite{vafa2020multi-defect}
\beq F^M = e^{i\psi}\frac{\pi}{4}\frac{1}{a} \zeta ,\eeq
i.e., a $+1/2$ defect travels at constant speed along its axis $e^{i\psi}$ (with opposite sign for the other defect, since for two defects, $e^{i\phi_j} = \pm e^{i\psi}$). The motile force $F^M$ has magnitude $\frac{\pi}{4} \frac{\zeta}{a}$ and makes an angle $\psi$ relative to the line connecting the defect to the apex (the radial line). Specifically, we can decompose $F^M$ into radial $F^M_r$ and tangential $F^M_\theta$ components as
\begin{subequations}
    \begin{align}
        F^M_r &= \frac{\pi}{4}\frac{\zeta}{a}\cos\psi \\
        F^M_\theta &= \frac{\pi}{4}\frac{\zeta}{a}\sin\psi
    \end{align}
\end{subequations}

Explicitly, for an active $+1/2$ defect labeled $1$ in the presence of another $+1/2$ defect labeled 2, the force is 
\begin{align}
    F_1^{\bar z} &= -\frac{\pi J \gamma^{-1}}{2} \left[\frac{1}{z_1 - z_2} + \frac{1}{z_1 - \tilde z_1} + \frac{1}{z_1 - \tilde z_2} -\frac{3}{2}\chi\frac{1}{z_1}\right]\nonumber \\
    &\qquad + e^{i\psi}\frac{\pi}{4}\frac{\zeta}{a} 
    \label{eq:ForcesDefectsCone}
\end{align}

In addition to the forces on defects, the Born-Oppenheimer approximation implies for a pair of neighboring defects that:
\begin{enumerate}
    \item the polarizations are anti-parallel~\cite{vromans2016orientational,vafa2020multi-defect}, and,
    \item the polarization relative to the line connecting the two defects does not change,
\end{enumerate}
\emph{even as the defects move}~\cite{vafa2022defectDynamics}.
In-line with all of our theoretical assumptions, we assess the validity of the Born-Oppenheimer approximation and its implications by directing comparing theoretical predictions with full numerical simulations.

\subsection{Simulation details}

\begin{figure}[t]
    \centering
    \includegraphics[width=\columnwidth]{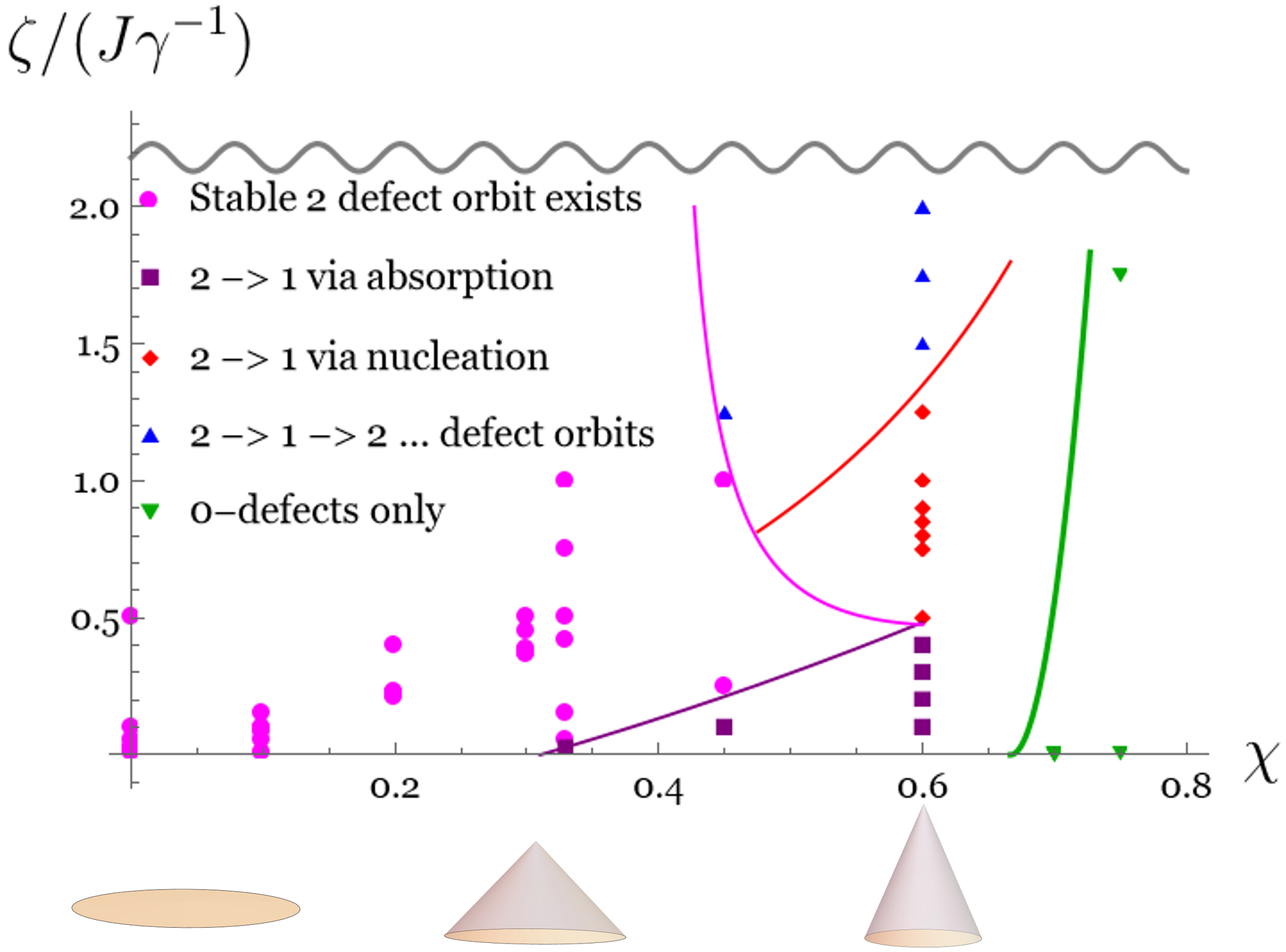}
    \caption{Phase diagram of main results with data from simulations included. The colored curves, based on the data, guide the eye, except for the red curve, which is given by Eq.~\eqref{eq:alpha_c} with $c'=0.9$.}
    \label{fig:phaseDiagramData}
\end{figure}
We check our theory with numerical simulations of the full active nematic texture on a cone, according to Eqs.~\eqref{eq:QDynamics}-\eqref{eq:num}. Without loss of generality, since a cone has zero Gaussian curvature everywhere except at the apex, we set $K'=0$ in the simulations. We impose strong anchoring boundary conditions at the base, i.e. the nematic director is at a fixed angle relative to the boundary. Explicitly, we impose at the boundary of the cone base of radius $R$ in the isothermal coordinate spaces of Fig.~\ref{fig:cone}(a), i.e. at $z = R e^{i\phi}$ where $\phi$ is the azimuthal coordinate):
\beq Q = e^{i\psi}e^{-\varphi + i \phi} \eeq
In our simulations, we solve Eq.~\eqref{eq:QDynamics} numerically using the method of lines~\cite{schiesser2012numerical}, where the temporal evolution is performed through a predictor-corrector scheme~\cite{press1992multistep} and spatial derivatives are evaluated using five-point stencil central differences. The phase diagram of our main results, which will be the central focus of the remainder of our paper, can be seen in Fig.~\ref{fig:phaseDiagramData}.

\section{Stable two defect orbits for $0 \le \chi < .31$}
\label{sec:circular}
We start with small activity on a disk and then extend to a cone by incrementally increasing the deficit angle $2\pi\chi$. We analyze two different set-ups in order: 1) two flank defects and then 2) one defect at the apex and the other on the flanks.

\subsection{Two defect orbits on a disk ($\chi = 0$)}

\begin{figure}[t]
    \centering
    \includegraphics[width=0.65\columnwidth]{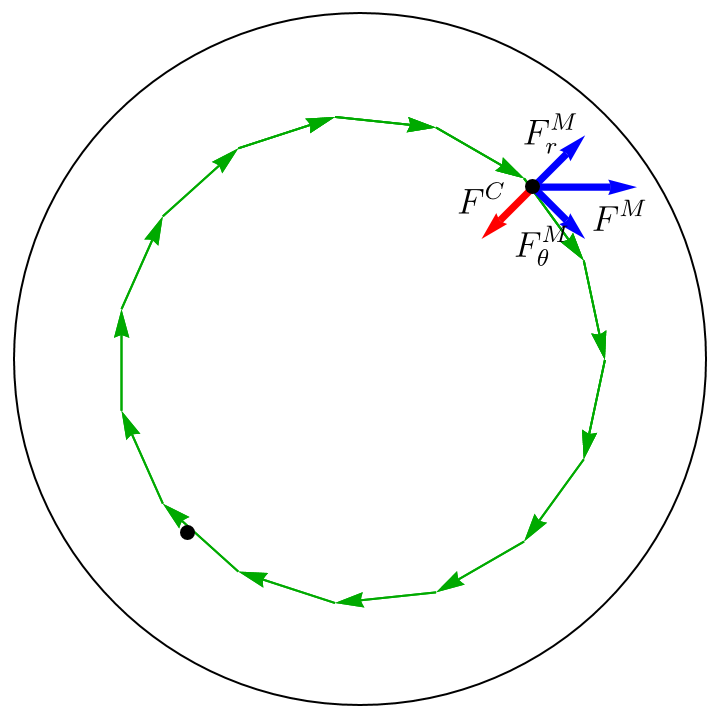}
    \caption{Sketch of forces for a pair of $+1/2$ defects on a disk.  Net Coulomb repulsion $F^C$ is in red, the motile force $F^M$, decomposed into its radial $F^M_r$ and tangential $F^M_\theta$ components, is in blue, and the defects follow the green arrows.}
    \label{fig:2DefectForcesDisk}
\end{figure}

\begin{figure*}
    \centering
    \subfloat[]{\includegraphics[width=.33\linewidth, trim={3cm 1cm 3cm 1cm},clip]{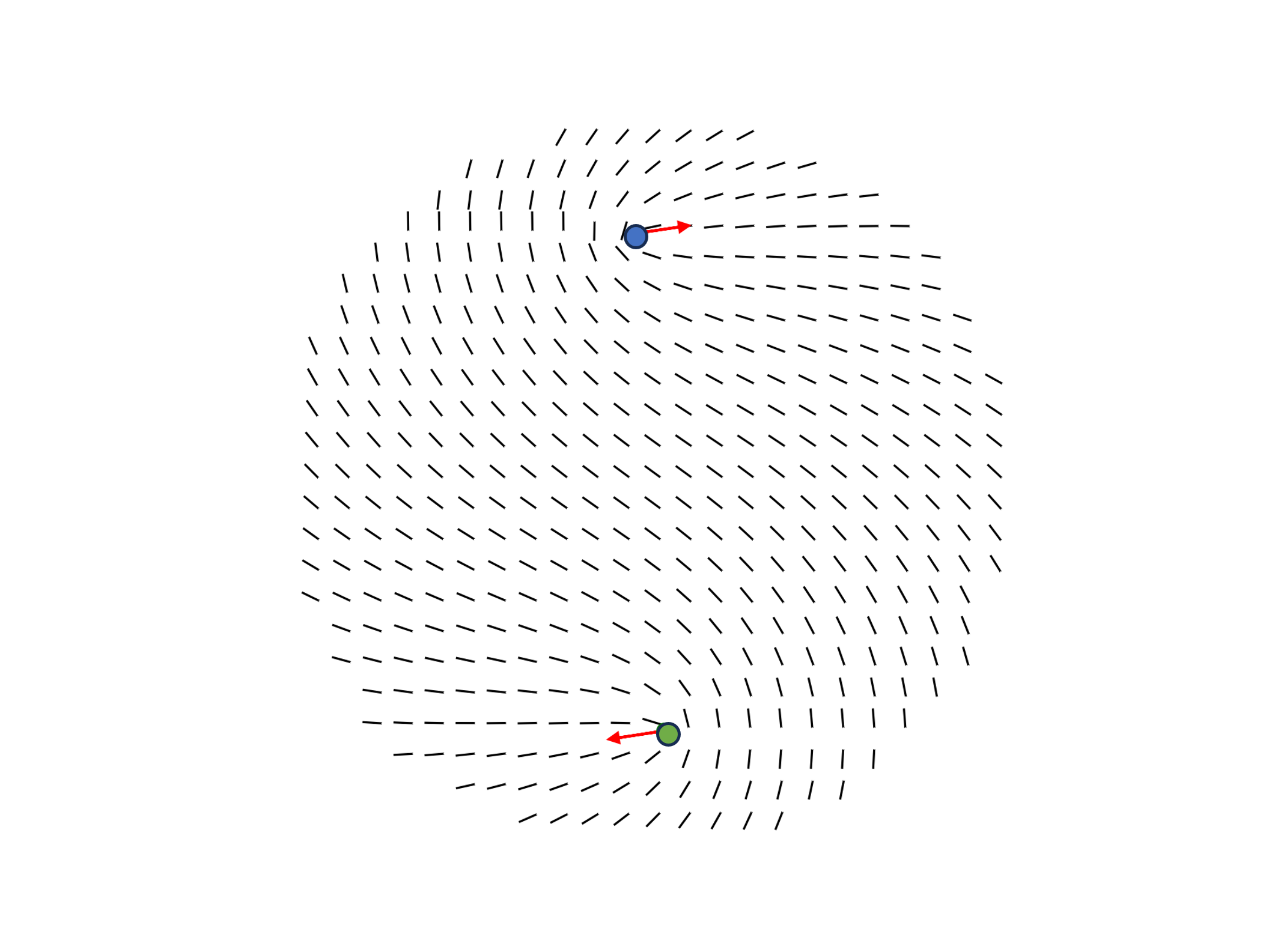}}
    \hfill
    \subfloat[]{\includegraphics[width=.33\linewidth, trim={3cm 1cm 3cm 1cm},clip]{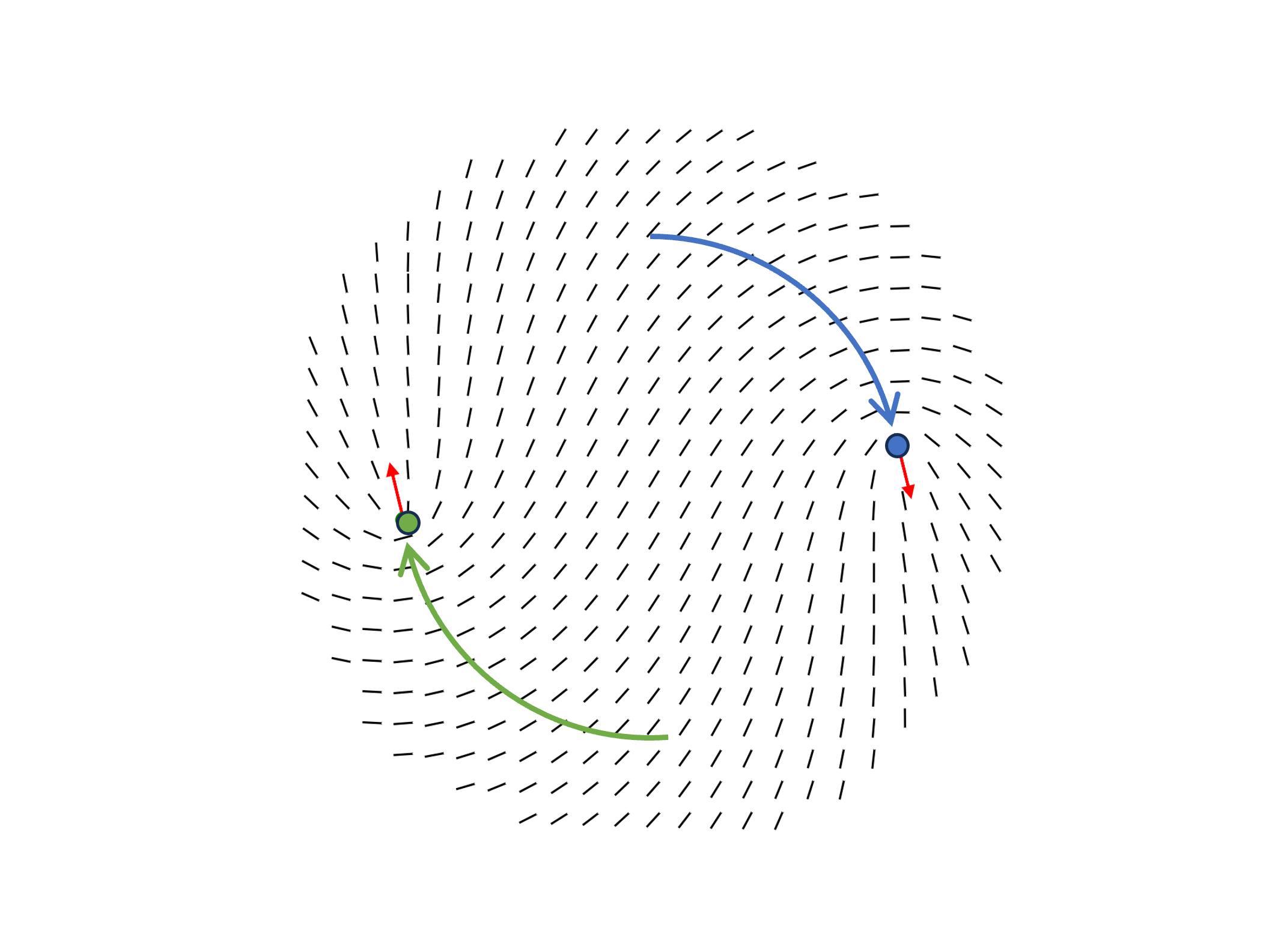}}
    \hfill
    \subfloat[]{\includegraphics[width=.33\linewidth, trim={3cm 1cm 3cm 1cm},clip]{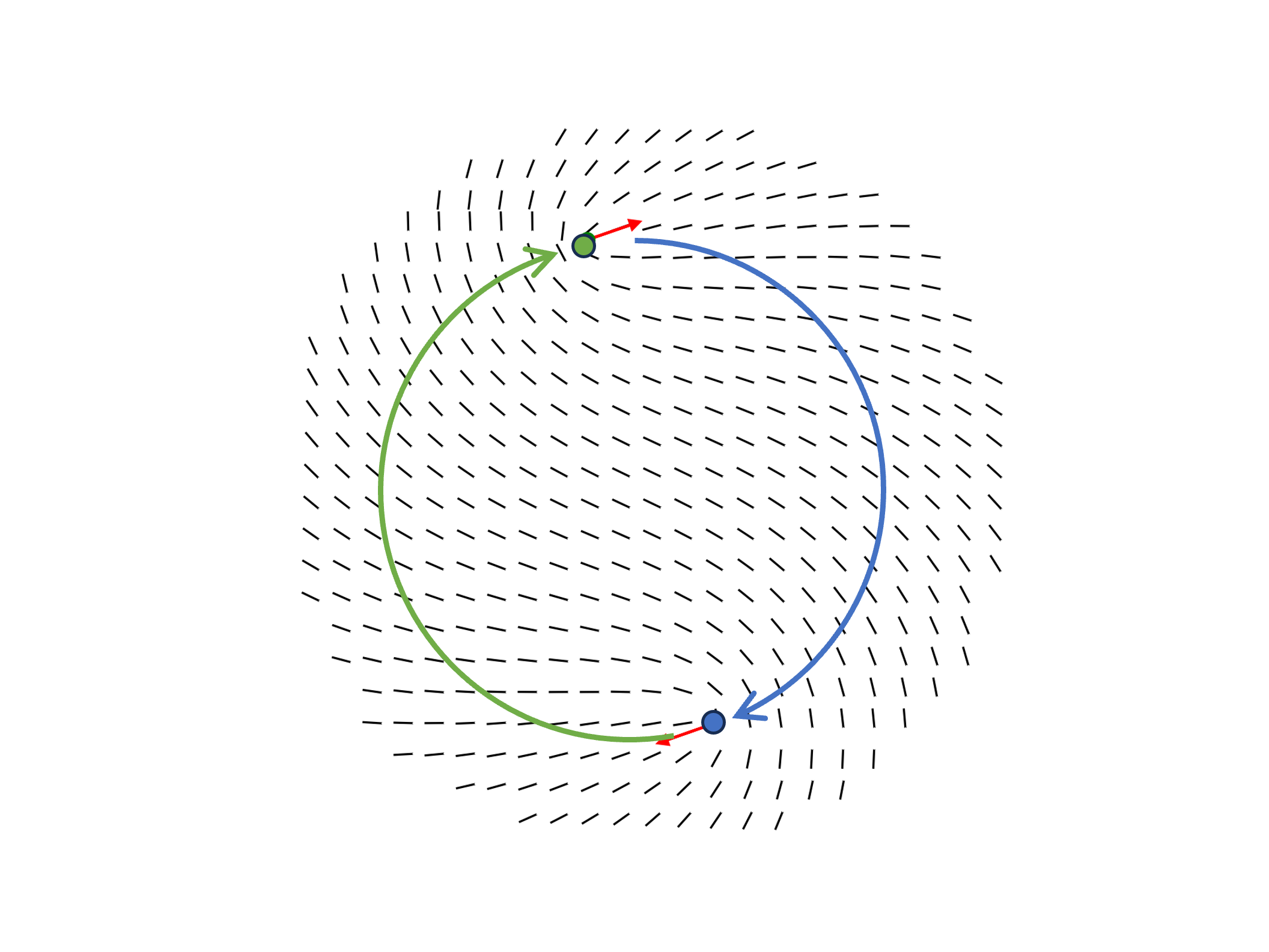}}
    \caption{Representative sequential snapshots of simulations on a disk for $\zeta/(J\gamma^{-1}) = 0.01$ and $\psi=\pi/2$. In all of the plots, the colored dots denote the $+1/2$ defects which follow the corresponding colored trajectories, and the red arrows denote the polarizations of the $+1/2$ defects.}
    \label{fig:disk}
\end{figure*}

To review, as discussed previously in Sec.~\ref{sec:stationary}, we know for a passive nematic on a disk that there is a classically stable two flank defect configuration~\cite{dzubiella2000topological,galanis2006spontaneous,duclos2017topological,vafa2022defectAbsorption}. Thus it is reasonable to expect that for small activity there should still be a classically stable two flank defect configuration, where now the defects can possibly move. We now show that there are indeed stable circular orbits on a disk.

We simulate an active nematic on a disk and we consider three cases: $\psi = 0, \pi/4, \pi/2$. In all three cases, in the simulations the distance of the defect to the center was constant. We now argue that this is consistent with the Born-Oppenheimer approximation (even in the case of the cone as well).

If the boundary phase $\psi = 0$, then since the motile force is radial, and the Coulomb force is always radial, all activity does is shift the equilibrium position, and hence we would still have static solution in this case. If the phase is $\psi = \pi/2$, since the motile force is perpendicular to the radial line, and remains perpendicular as the defect moves as explained in the previous section, the distance from the apex does not change, and the defect moves in a circular orbit. For generic phase $\psi$, the radial distance readjusts so that the net Coulomb force (including any contribution from the cone apex) balances the radial component of the motile force, and due to the tangential component of the motile force, the flank defect undergoes circular motion.

Explicitly, we solve for the distance $r$ by setting the radial component of the net force given in Eq.~\eqref{eq:ForcesDefectsCone} to zero. Since for a disk $\chi=0$, we find that
\beq \frac{\pi J \gamma^{-1}}{2}\left[\frac{1}{2 r} + \frac{1}{R^2/r + r} - \frac{1}{R^2/r - r}\right] = \frac{\pi}{4}\frac{\zeta}{a}\cos\psi .\eeq	

See Fig.~\ref{fig:2DefectForcesDisk} for a sketch. What this means is that generically defects will have circular orbits with constant speed due to vanishing net radial force and non-zero azimuthal component of the motile force. We want to emphasize that unlike the mechanism of gravity, here there is no centripetal force, since the net radial force vanishes.

For $\psi = 0$, as expected, there was no orbit, since the motility has no tangential component. For $\psi = \pi/4, \pi/2$, defects formed circular orbits (see Fig.~\ref{fig:disk} for representative snapshots of stable circular orbits of two defects on a disk for $\psi=\pi/2$). For all 3 values of $\psi$, the radial components of the defects, with defect core size $a=0.9$, was as expected. Moreover, using the mobility matrix from Ref.~\cite{vafa2020multi-defect},
\beq \mathcal M_{ii} \dot z_i = -\frac{\pi J \gamma^{-1}}{2}\left[\frac{1}{2 r} + \frac{1}{R^2/r + r} - \frac{1}{R^2/r - r}\right] + \frac{\pi}{4}\frac{\zeta}{a}\cos\psi \eeq
where
\begin{align}
    \mathcal M_{ii} &= \frac{\pi}{2} \sigma_i^2
    \left[\ln \left(\frac{R^2 - r_i^2}{a^2}\right) - \frac{R^2 - 2r_i^2}{R^2 - r_i^2}\right.\nonumber \\
    &{} \left. + \frac{1}{2}\frac{R^4 \left(r_i^2 \left(3 r_i^2-2 R^2\right)-2 \left(R^2-r_i^2\right)^2 \ln \left(1- \frac{r_i^2}{R^2}\right)\right)}{r_i^6 \left(r_i^2-R^2\right)}\right]
\end{align}
the theoretical prediction for the speed was confirmed by the simulations. There is surprisingly good agreement given that we know that the ansatz is not exact. Assuming the mobility is independent of the defect position, we find theoretically that the ratio of speeds is $\sqrt{2} \approx 1.4$, which is close to the numerical value of $(80/89)/(88/146) \approx 1.5$. In Table~\ref{tab:disk}, we provide a quantitative comparison between the numerical results and theoretical predictions by including the mobility, showing an excellent agreement and verifying the validity of our Born-Oppenheimer approximation.

\begin{table*}[t]
    \begin{tabular}{ | c | c | c | c | c | }
        \hline 
        Phase $\psi$ & Theoretical distance & Numerical distance&  Theoretical speed ($\zeta/a$) & Numerical speed ($\zeta/a$) \\
        \hline \hline
        $0$   & 0.76R & 0.76R & 0  & 0 \\
        \hline
        $\pi/4$   & 0.75R & 0.75R & $1.9 \times 10^{-4}$  & $1.9 \times 10^{-4}$ \\
        \hline
        $\pi/2$  &0.68R &0.68R & $2.7 \times 10^{-4}$   & $2.8 \times 10^{-4}$ \\
        \hline
    \end{tabular}
    \caption{Table of distances and speeds for $\zeta/(J\gamma^{-1}) = 0.01$ and $\psi = 0, \pi/4, \pi/2$ for both theoretical prediction and simulation result on a disk. Here $a = 0.9$.}
    \label{tab:disk}
\end{table*}

\subsection{Two defect orbits on cone with $0 < \chi < .31$}

\begin{figure}[t]
    \centering
    \includegraphics[width=0.65\columnwidth]{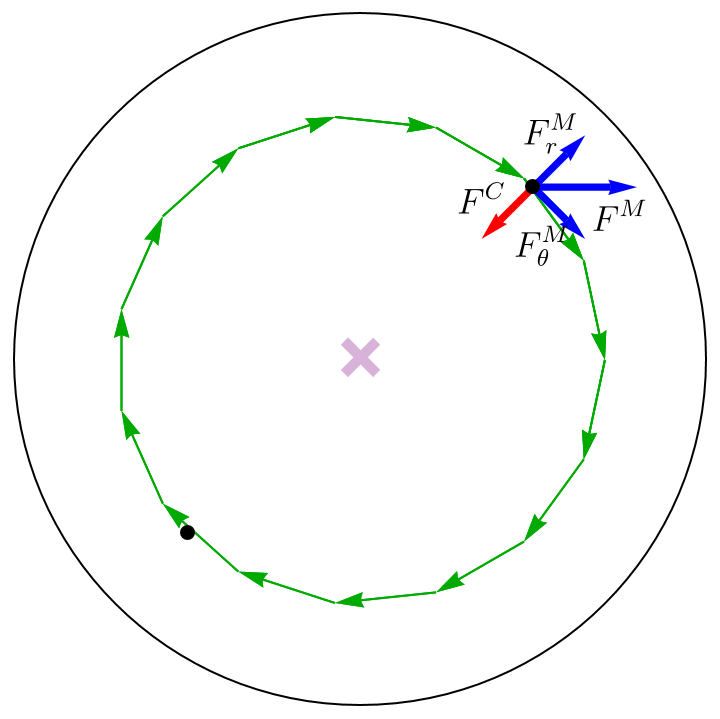}
    \caption{Sketch of forces for a pair of $+1/2$ defects on a cone. Black dots denote $+1/2$ defects and purple star denotes apex. Net Coulomb repulsion $F^C$ is in red, the motile force $F^M$, decomposed into its radial $F^M_r$ and tangential $F^M_\theta$ components, is in blue, and the defects follow the green arrows.}
    \label{fig:2DefectForcesCone}
\end{figure}

Having verified the dynamics against full simulations on a disk, we now begin with extending our analyses to the cone, incrementally increasing the deficit angle $2\pi\chi$.
For $\zeta=0$ and $\chi < 0.31$, we know that there is a classically stable two flank defect configuration, as just previously discussed. For small deficit angles, similar to the case of a disk, it is reasonable to expect that for small activity there should still be a classically stable two flank defect configuration, where the defects can possibly move. We now show that for small activity there are stable circular orbits on a cone for $\chi < 0.31$.

We again solve for the distance $r$ by setting the radial component of the net force given in Eq.~\eqref{eq:ForcesDefectsCone} to zero, leading to
\beq \frac{\pi J \gamma^{-1}}{2}\left[-\frac{3}{2} \frac{\chi}{1-\chi}\frac{1}{r} + \frac{1}{2 r} + \frac{1}{R^2/r + r} - \frac{1}{R^2/r - r}\right] = \frac{\pi}{4}\frac{\zeta}{a}\cos\psi \label{eq:ForceBalance2DefectCone} \eeq	

See Fig.~\ref{fig:2DefectForcesCone} for a sketch. As before, since the radial component of the motile force, if any, balances the net Coulomb force, and the tangential component of the motile force is constant, then generically defects will have circular orbits with constant speed, with a $\chi$-dependent radius. The behavior is thus qualitatively independent of $\chi$. See Table~\ref{tab:twodefectschip1} for comparison of theory to simulations for $\chi = 1/10$, $\frac{\zeta}{J \gamma^{-1}} = .01$, and $\psi = 0, \pi/4, \pi/2$.

\begin{table}[t]
    \begin{tabular}{ | c | c | c| }
        \hline 
        Phase $\psi$ & Theoretical distance & Numerical distance \\
        \hline \hline
        $0$   & 0.70 R & 0.70 R \\
        $\pi/4$  & 0.68 R & 0.69 R \\
        \hline
        $\pi/2$  &0.62 R & 0.63 R \\
        \hline
    \end{tabular}
    \caption{Table of distances for $\chi=1/10$ and $\zeta/(J \gamma^{-1}) = .01$ for $\psi = 0, \pi/4, \pi/2$ for both theoretical prediction and simulation results. Here $a = 0.9$.}
    \label{tab:twodefectschip1}
\end{table}

\section{Single defect orbits and defect unbinding on a cone}
\label{sec:oneDefect}
As shown in sec.~\ref{sec:stationary} for a passive nematic texture on a cone, the global minimum of the configuration is a single $+1/2$ defect on the cone flank. This ground state provides a peculiar setup to investigate basic features of an isolated $+1/2$ defect on a flank and its active dynamics.

\subsection{Single defect orbits}

\begin{figure}[t]
    \centering
    \includegraphics[width=0.65\columnwidth]{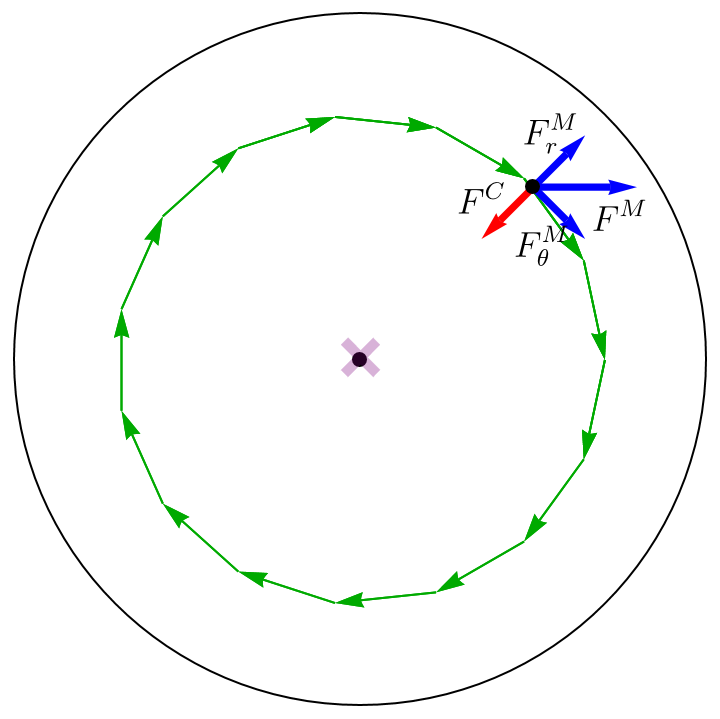}
    \caption{Sketch of forces for a pair of $+1/2$ defects on a cone, where one defect is at the apex and the other is on the flanks. Black dots denote $+1/2$ defects and purple star denotes apex. Net Coulomb repulsion $F^C$ is in red, the motile force $F^M$, decomposed into its radial $F^M_r$ and tangential $F^M_\theta$ components, is in blue, and the flanks defect follows the green arrows.}
    \label{fig:1DefectForcesCone}
\end{figure}

\begin{table}[t]
    \begin{tabular}{ | c | c | c| }
        \hline 
        Phase $\psi$ & Theoretical distance & Numerical distance \\
        \hline \hline
        $\pi/4$   & 0.72R & 0.73R \\
        \hline
        $\pi/2$  &0.71R & 0.69R \\
        \hline
    \end{tabular}
    \caption{Table of distances for $\chi=1/10$ and $\zeta/(J \gamma^{-1}) = .01$ for $\psi = \pi/4, \pi/2$ for both theoretical prediction and simulation result on a cone. Here $a = 0.9$.}
    \label{tab:1defectchip1}
\end{table}

\begin{table}[t]
    \begin{tabular}{ | c | c | c| }
        \hline 
        Phase $\psi$ & Theoretical distance & Numerical distance \\
        \hline \hline
        $0$   & 0.88R & 0.87R \\
        \hline
        $\pi/2$  &0.58R & 0.78R \\
        \hline
    \end{tabular}
    \caption{Table of distances for $\chi=1/3$ and $\zeta/(J \gamma^{-1}) = 0.25$ for $\psi = 0, \pi/2$ for both theoretical prediction and simulation result on a cone. Here $a=0.9$.}
    \label{tab:1defectchip33}
\end{table}

We start with the single flanks defect ground state solution obtained in \cite{vafa2022defectAbsorption}, which is valid for $\chi < 2/3$. We now add small activity. 

As usual, we solve for the distance $r$ of the isolated flank defect from the apex by  setting the radial component of the net force given in Eq.~\eqref{eq:ForcesDefectsCone} to zero, leading to
\beq \frac{\pi J \gamma^{-1}}{2}\left[-\frac{3}{2} \frac{\chi}{1-\chi}\frac{1}{r} + \frac{1}{r} - \frac{1}{R^2/r - r}\right] = \frac{\pi}{4}\frac{\zeta}{a}\cos\psi \eeq	

See Fig.~\ref{fig:1DefectForcesCone} for a sketch. Similar to the two-defect case, the defect will circularly orbit the apex at constant speed, which again unlike gravity is not driven by a centripetal force, since the net radial force vanishes.

For comparison of theory to numerics, see Table~\ref{tab:1defectchip1}  for $\chi=1/10$, $\frac{\zeta}{J \gamma^{-1}} = 0.01$, and $\psi = \pi/4, \pi/2$, and Table~\ref{tab:1defectchip33} for $\chi=1/3$, $\frac{\zeta}{J \gamma^{-1}} = 0.25$, and $\psi = 0, \pi/2$. These tables suggest that for small $\chi$ and $\zeta$, the multi-defect ansatz is good, but becomes worse as $\chi$ and $\zeta$ increase. We observed in the simulations that the polarizations of the defects rotated to point outwards, consistent with the observation that the numerical distance is larger than the prediction in Table~\ref{tab:1defectchip33}. This discrepancy implies that the global phase $\psi$ becomes space-dependent as $\chi$ and $\zeta$ increase, which is not captured in our ansatz. 

\subsection{$1 \to 2$ defect transition via defect unbinding}

\begin{figure*}
    \centering
    \subfloat[]{\includegraphics[width=.25\linewidth, trim={3cm 1cm 3cm 1cm},clip]{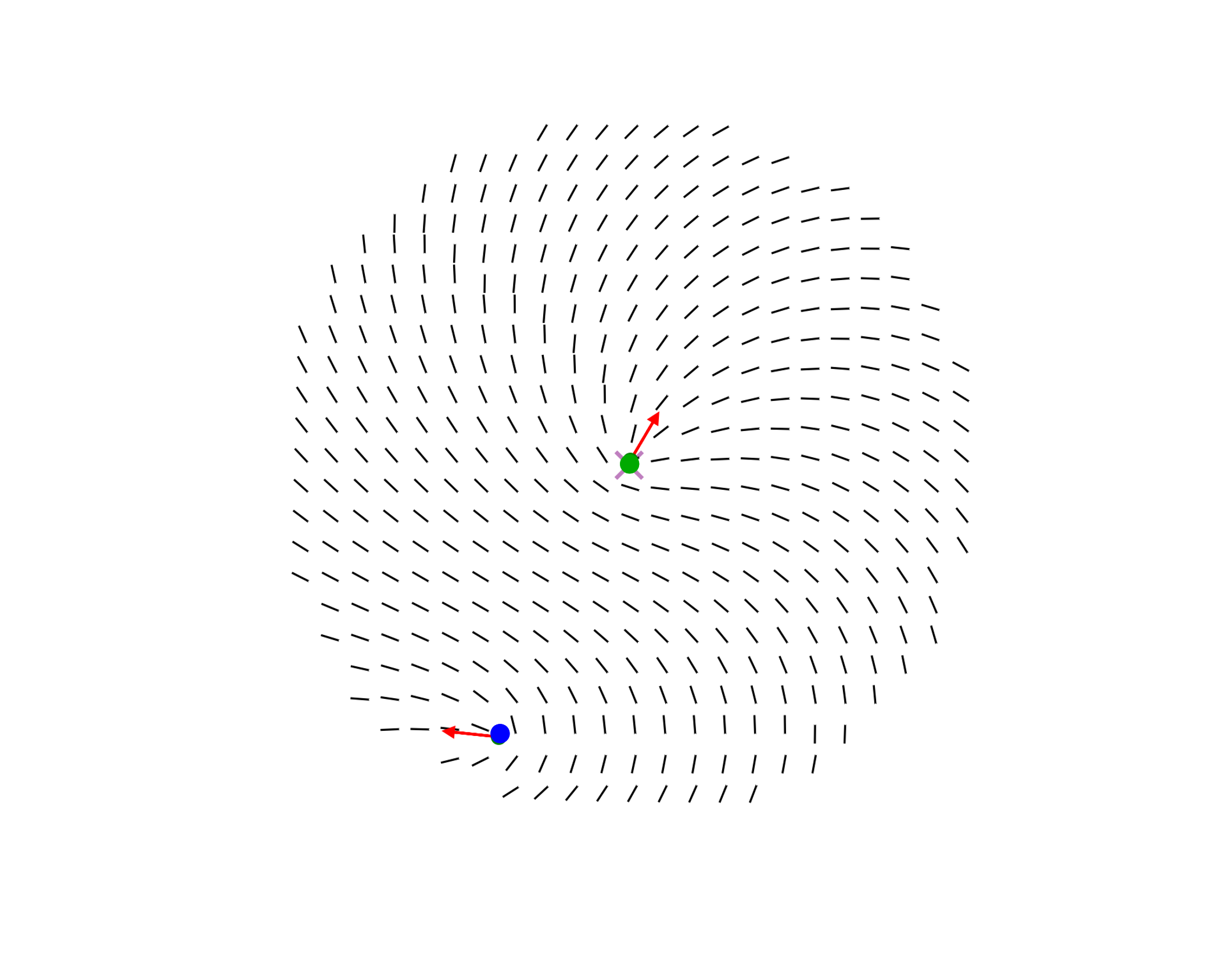}}
    \hfill
    \subfloat[]{\includegraphics[width=.25\linewidth, trim={3cm 1cm 3cm 1cm},clip]{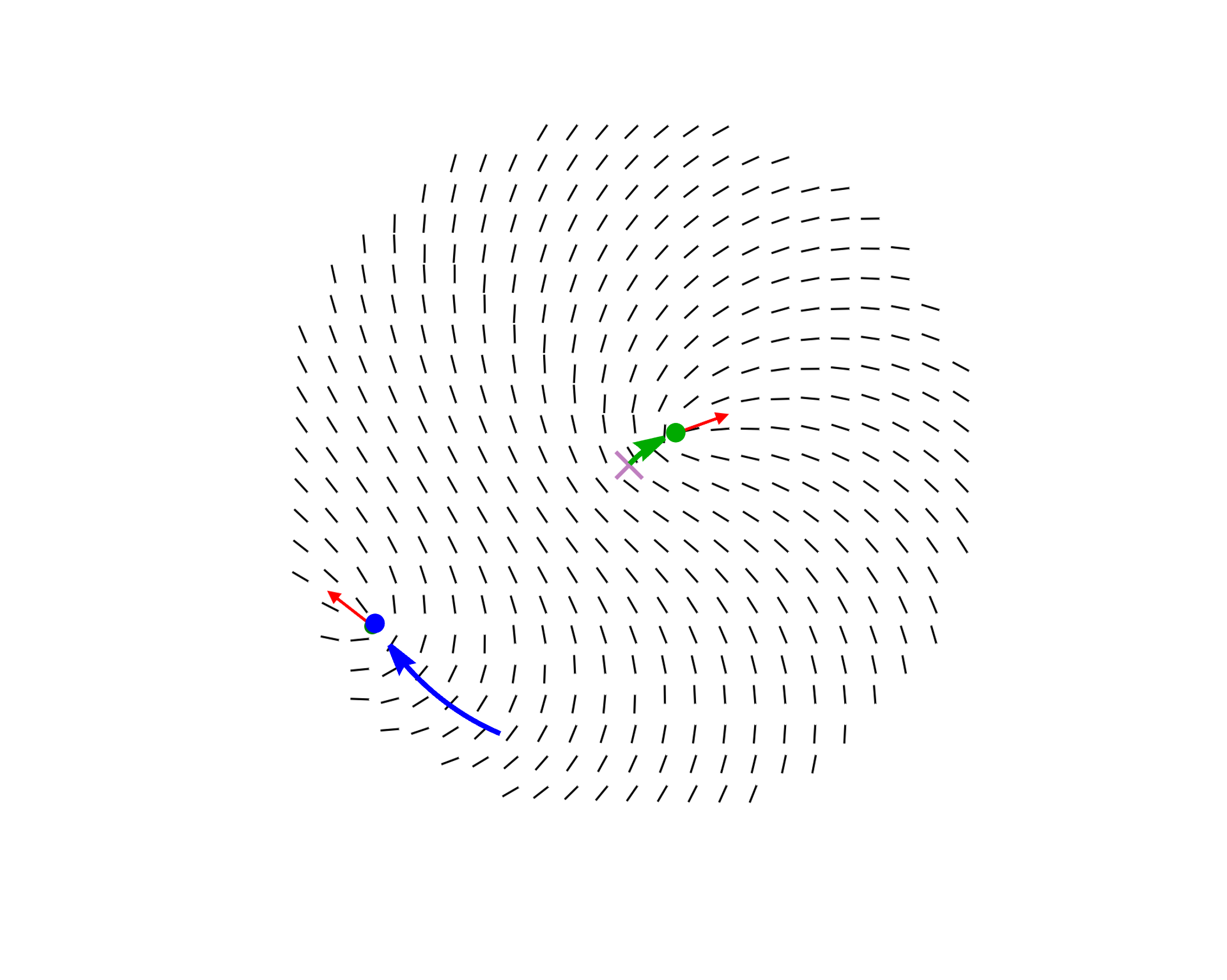}}
    \hfill
    \subfloat[]{\includegraphics[width=.25\linewidth, trim={3cm 1cm 3cm 1cm},clip]{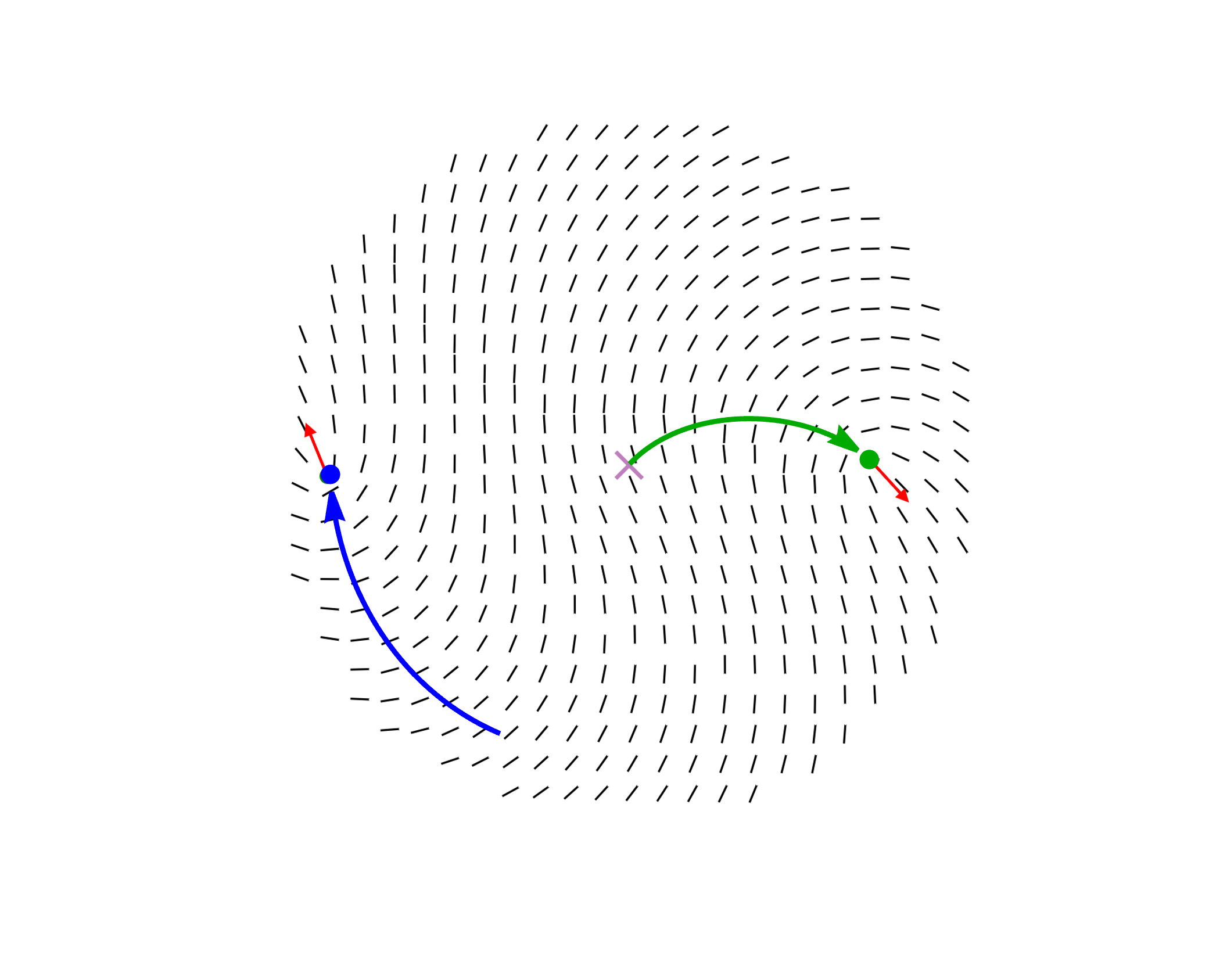}}
    \hfill
    \subfloat[]{\includegraphics[width=.25\linewidth, trim={3cm 1cm 3cm 1cm},clip]{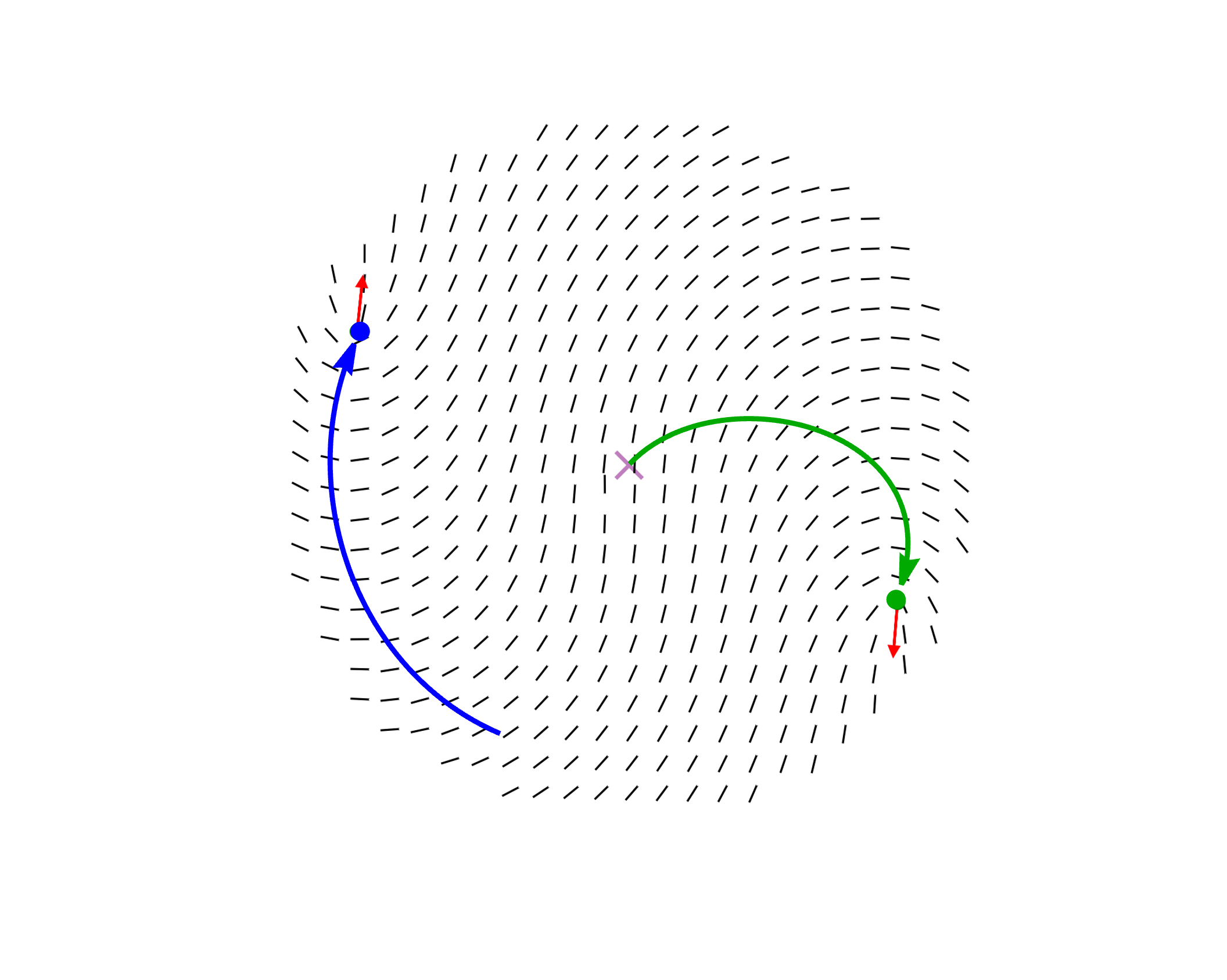}}\\
    \subfloat[]{\includegraphics[width=.25\linewidth]{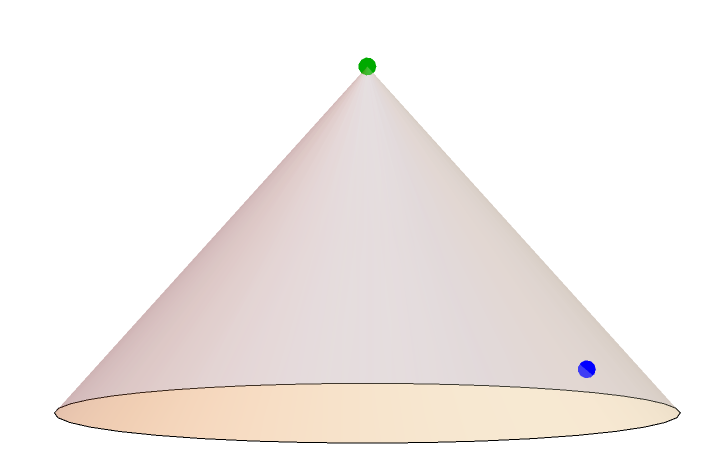}}
    \hfill
    \subfloat[]{\includegraphics[width=.25\linewidth]{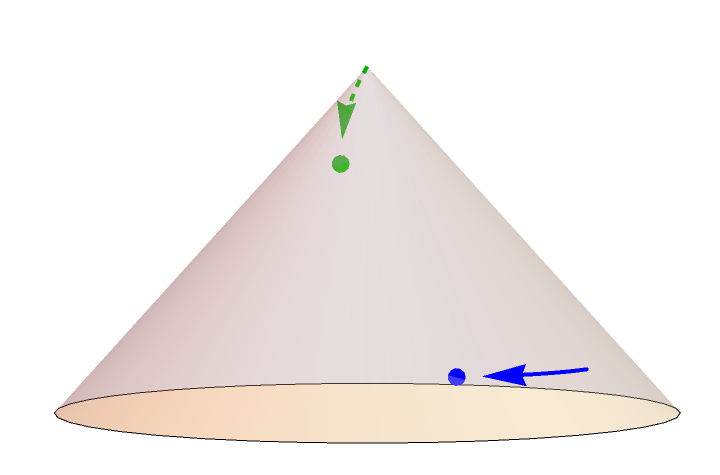}}
    \hfill
    \subfloat[]{\includegraphics[width=.25\linewidth]{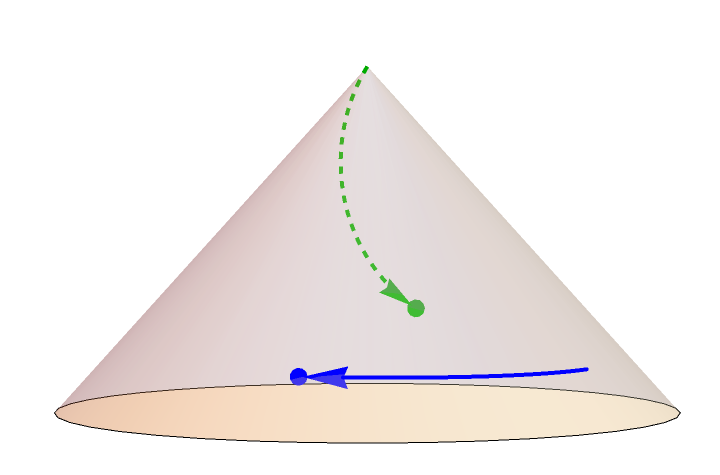}}
    \hfill
    \subfloat[]{\includegraphics[width=.25\linewidth]{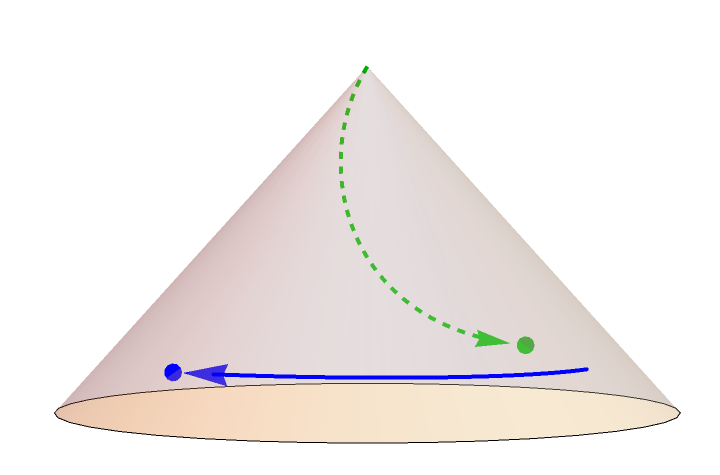}}
    \caption{Key snapshots of simulations depicting the defect unbinding from the apex mechanism. Top row: plots are in isothermal coordinates on a cone with $\chi=0.33$ and $\zeta/(J\gamma^{-1}) = 0.5$ and $\psi=\pi/2$. (a) Initially a single defect is orbiting. (b) the $+1/2$ defect begins unbinding from the apex. (c) the $+1/2$ defect has successfully unbound (escaped) from the apex. (d) representative snapshot of stable circular two defect orbit. In all of the plots, the colored dots denote the $+1/2$ defects which follow the corresponding colored trajectories, the red arrows the polarizations of the $+1/2$ defects, and the purple cross at the origin denotes the apex. In the bottom row are the corresponding plots on a 3D cone.}
    \label{fig:simulation_unbinding}
\end{figure*}
\begin{figure}[t]
    \centering
    \includegraphics[width=\columnwidth]{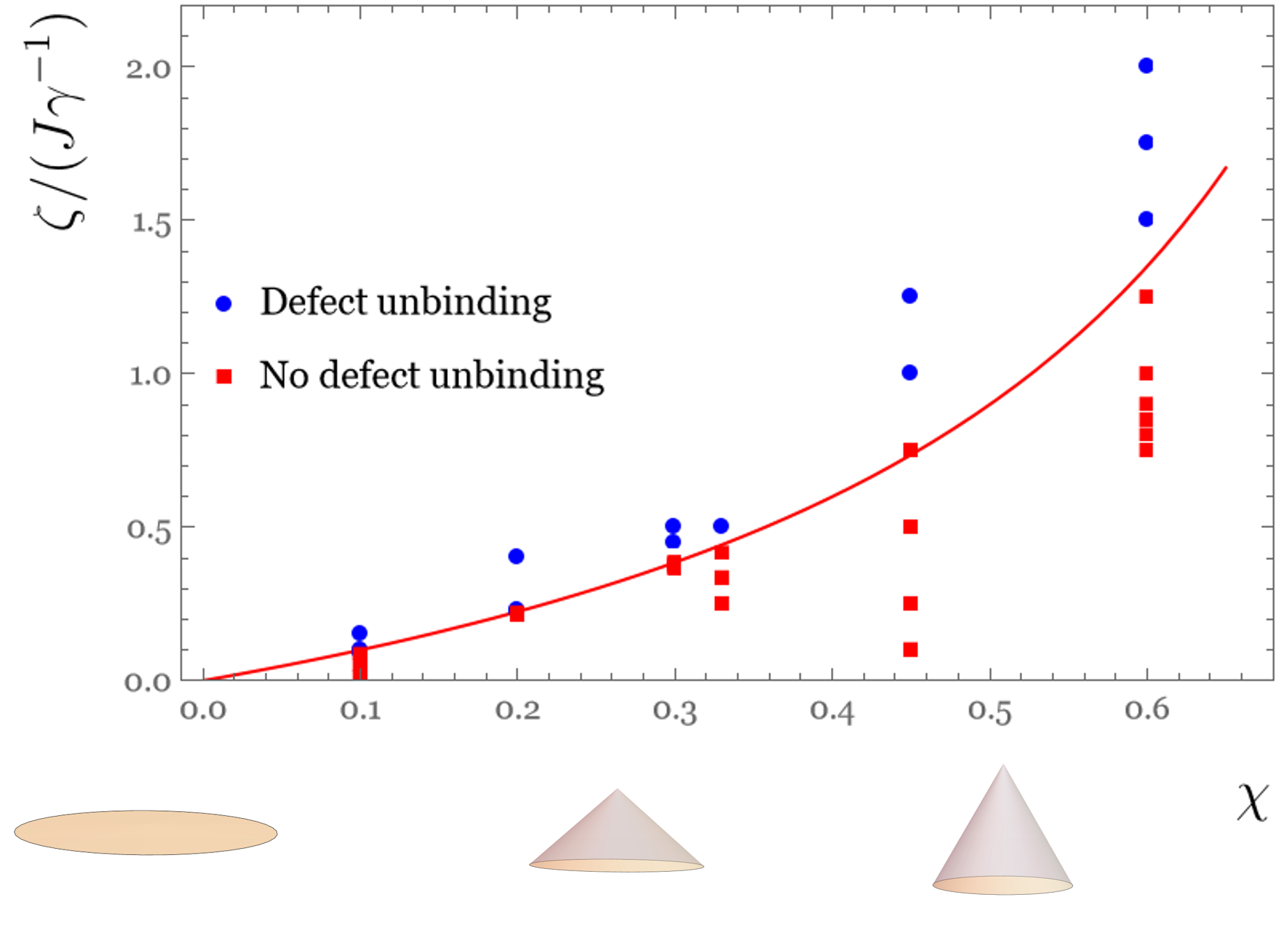}
    \caption{Plot of data from simulations for when defect unbinding occurs. Red curve corresponds to the analytical prediction (Eq.~\eqref{eq:alpha_c} with $c'=0.9$).}
    \label{fig:defectUnbinding}
\end{figure}

Numerical simulations allow us to explore a larger phase space of stronger activities. While 
for a small activity, in accordance with our theoretical prediction, we observed one defect at the apex and a single defect orbit on the cone flank, for sufficiently large activity, the $+1/2$ defect at the apex experiences a stronger motile force, which can be sufficient to allow the defect to escape the Coulombic attraction to the apex due to the deficit angle. As the defect unbinds from the apex, the defects on the flanks can experience circular motion equidistant from the apex as before (see Fig.~\ref{fig:simulation_unbinding} for key snapshots depicting the unbinding mechanism for $\psi=\pi/2$.)

We find the critical value for the activity $\zeta_c$ by equating the motile force with the Coulomb force at the apex in physical coordinates~\cite{vafa2023active}:
\beq 3c \frac{\chi}{1-\chi}\frac{ J\gamma^{-1}}{a} = \frac{\zeta_c}{a} \implies \zeta_c = c'J\gamma^{-1}\frac{\chi}{1-\chi} , \label{eq:alpha_c}\eeq
where $c' = 3c$. We introduced a factor of $c$ in the LHS to incorporate the defect core size $\sim a$, and from simulations learn that $c' \approx 0.9$. Remarkably, comparison of this analytically calculated critical activity for defect unbinding with the results of full numerical simulations shows excellent agreement (see Fig.~\ref{fig:defectUnbinding}). As such, we argue that for any experimental realization of an active nematic on a cone, using Eq.~\ref{eq:alpha_c}, and knowing the material constants of the nematic texture, i.e. the Frank elasticity $J$ and orientational diffusion $\gamma$, the critical activity for the unbinding of nematic topological defects can be found for any conical geometry with deficit angle $2\pi\chi$.

\section{Large deficit angle / activity}
\label{sec:large}

\subsection{Stable two defect configurations for $\chi > 0.31$}

\begin{table}[t]
    \begin{tabular}{ | c | c | c| }
        \hline 
        Phase $\psi$ & Theoretical distance & Numerical distance \\
        \hline \hline
        $0$   & 0.93 R$^*$ & 0.93 R \\
        \hline
        $\pi/2$  & N/A & 0.86 R \\
        \hline
    \end{tabular}
    \caption{Table of distances for $\chi=1/3$ and $\zeta/(J \gamma^{-1}) = .5$ for $\psi = 0, \pi/2$ for both naive theoretical prediction and simulation result on a cone. Here $a = 0.9$.}
    \label{tab:twodefectschip33}
\end{table}

For $\chi > 0.31$ and sufficiently small activity, we expect there to be no stable two flanks defect configuration, so if we start with two flank defects, we should end up with 1 flank defect and the other defect absorbed by the apex. We recently analyzed defect absorption by the apex in Ref.~\cite{vafa2023active}, and from simulations we do indeed see defect absorption.

However, from simulations we learn that sufficiently large activity can stabilize the two defect orbit. See Table~\ref{tab:twodefectschip33} for comparison of theory to numerics for $\chi=1/3$, $\frac{\zeta}{J \gamma^{-1}} = .5$, and $\psi = 0, \pi/2$. Although we do not have an explanation as to why a two defect configuration is stable for $\psi=0$, naively using force balance ( Eq.~\eqref{eq:ForceBalance2DefectCone}) (assuming stability) surprisingly leads to a distance that is consistent with the numerical observation. Moreover, for $\psi = \pi/2$, since again we do not have an argument for a stable two defect orbit, which was observed in our simulations, then a possible explanation is that the defect polarizations rotated to point outwards, counteracting the attraction to the cone due to the deficit angle. We indeed observe such behavior, further supporting the observation that as $\chi$ and $\zeta$ increase, the multi-defect ansatz becomes worse. As with the case of single defect orbit on a cone with $\chi = 1/3$, we observed in the simulations for $\psi = \pi/2$ that the polarizations of the defects rotated to point outwards, consistent with the observation that the numerical distance is larger than the theoretical prediction. This discrepancy implies that the global phase $\psi$ becomes space-dependent as $\chi$ and $\zeta$ increase, which is not captured in our ansatz.

\subsection{Apex-mediated defect nucleation and emission}
\label{subsec:nucleation}

\begin{figure*}[t]
    \centering
    \subfloat[]{\includegraphics[width=.25\linewidth, trim={3cm 1cm 3cm 1cm},clip]{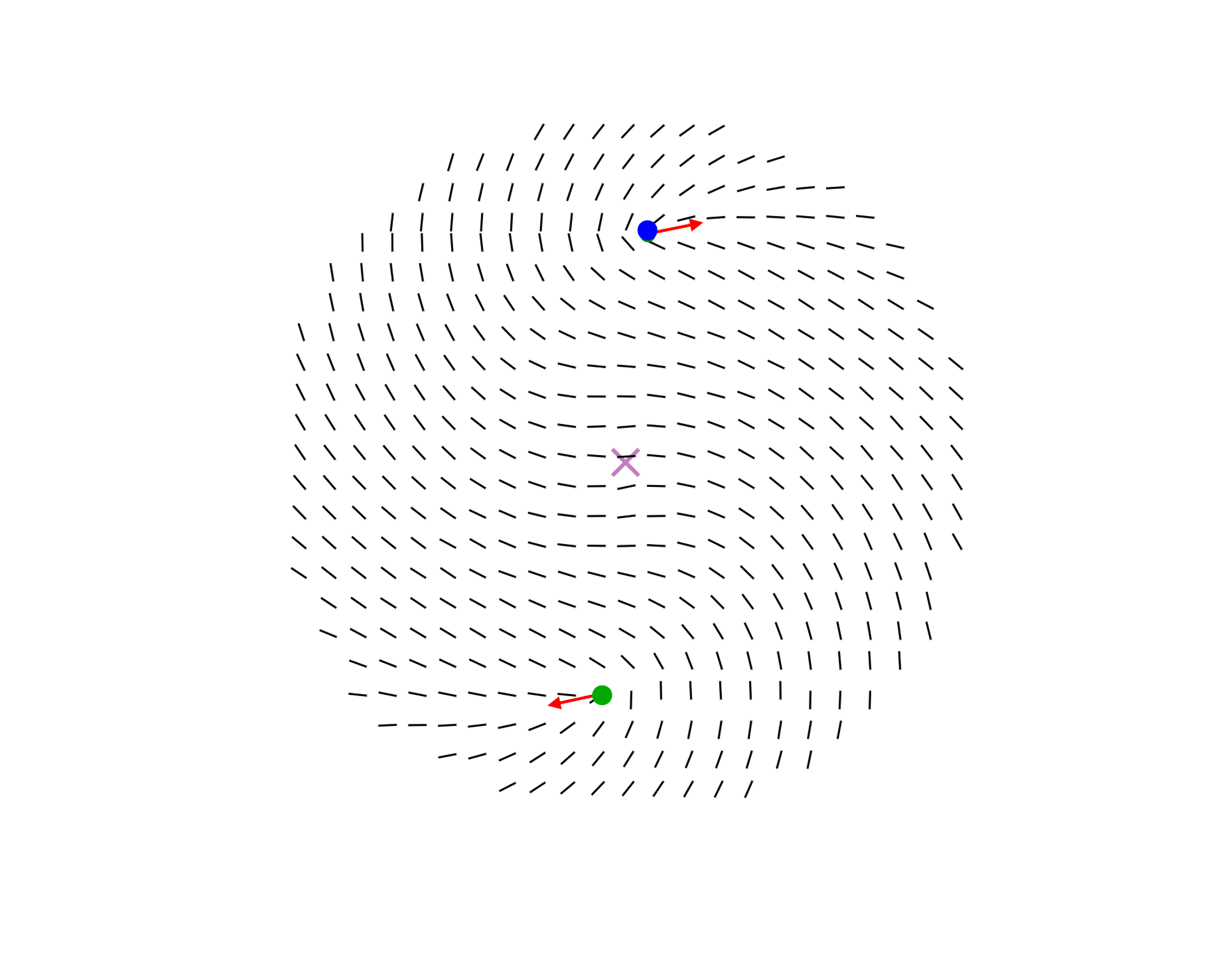}}
    \hfill
    \subfloat[]{\includegraphics[width=.25\linewidth, trim={3cm 1cm 3cm 1cm},clip]{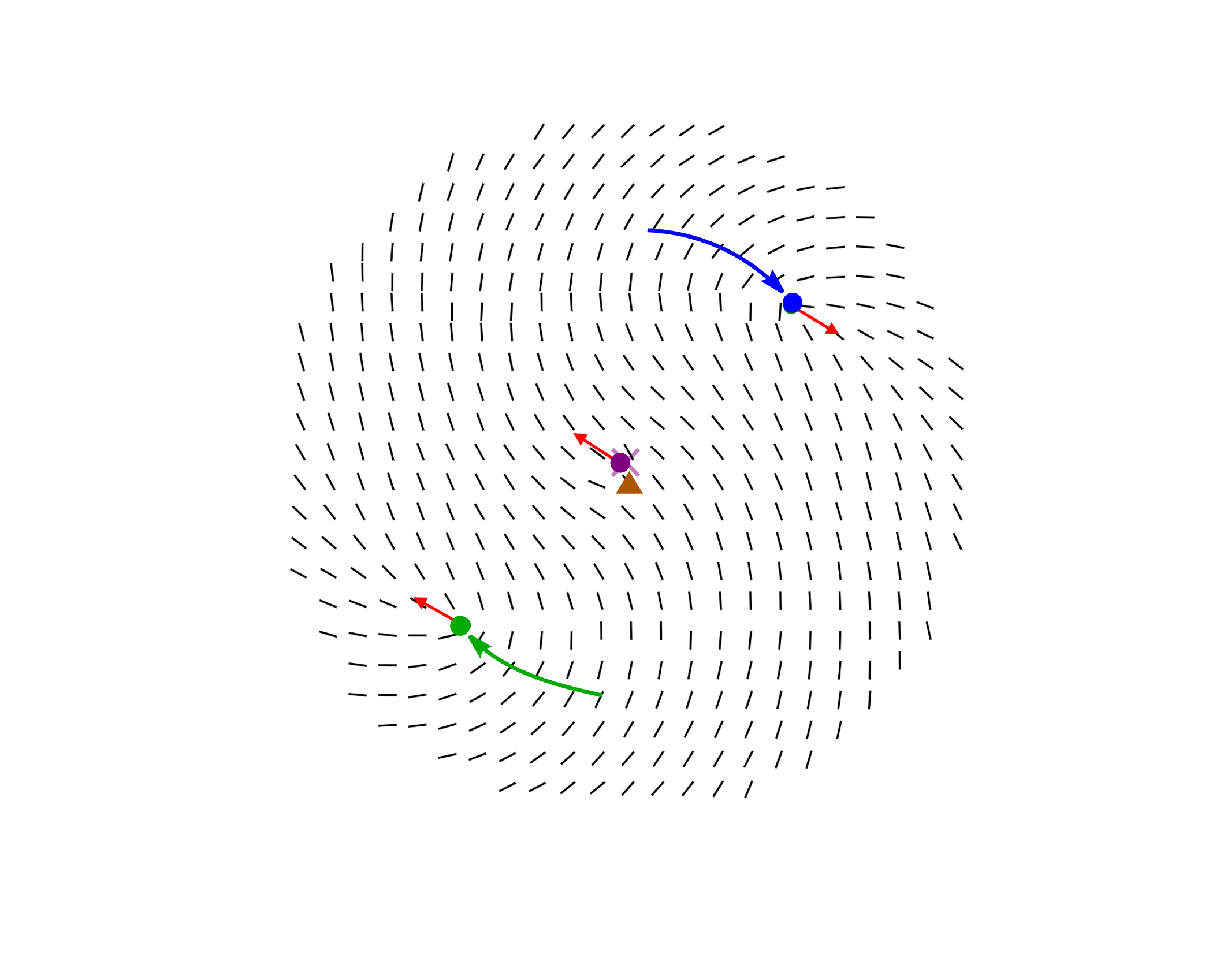}}
    \hfill
    \subfloat[]{\includegraphics[width=.25\linewidth, trim={3cm 1cm 3cm 1cm},clip]{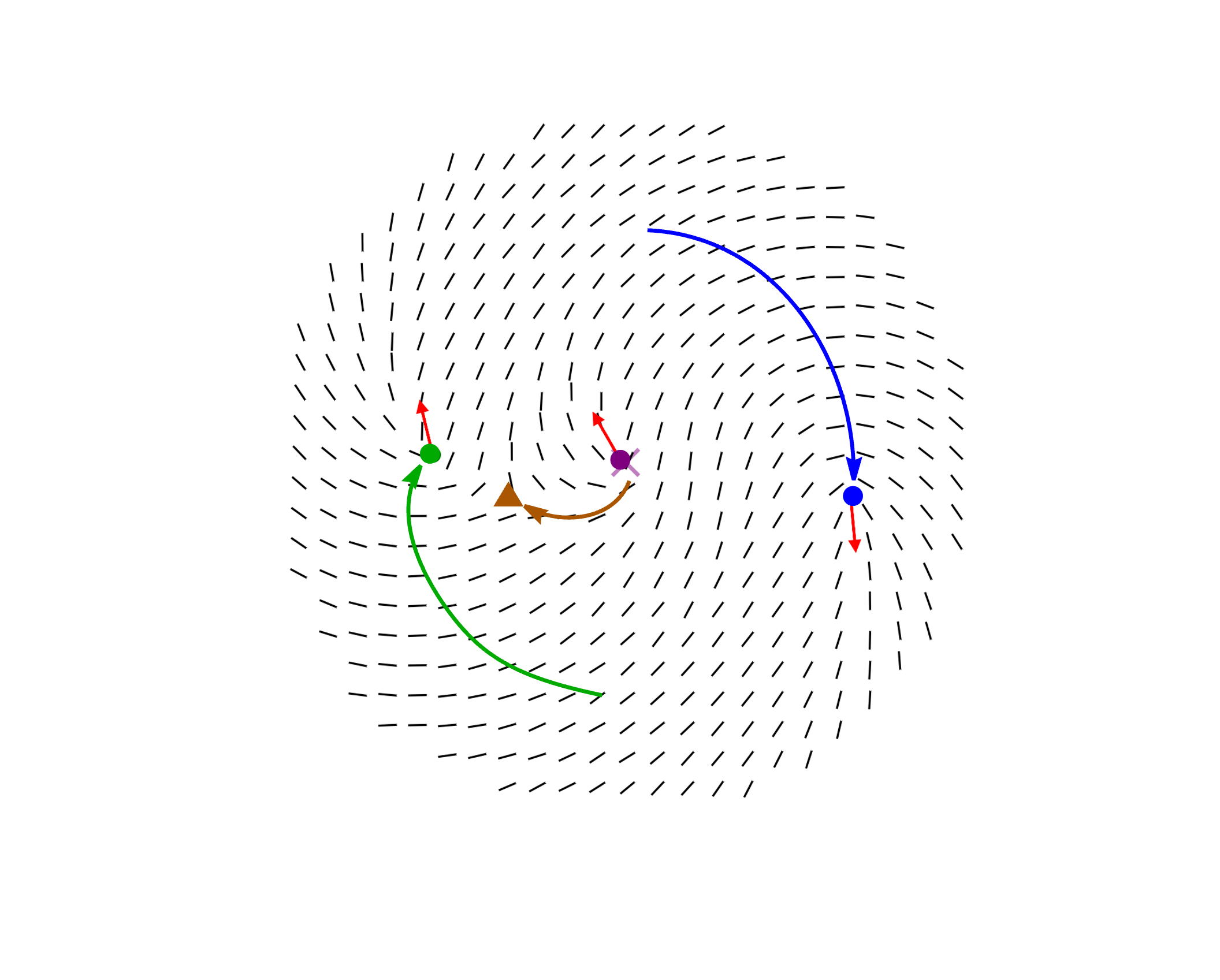}}
    \hfill
    \subfloat[]{\includegraphics[width=.25\linewidth, trim={3cm 1cm 3cm 1cm},clip]{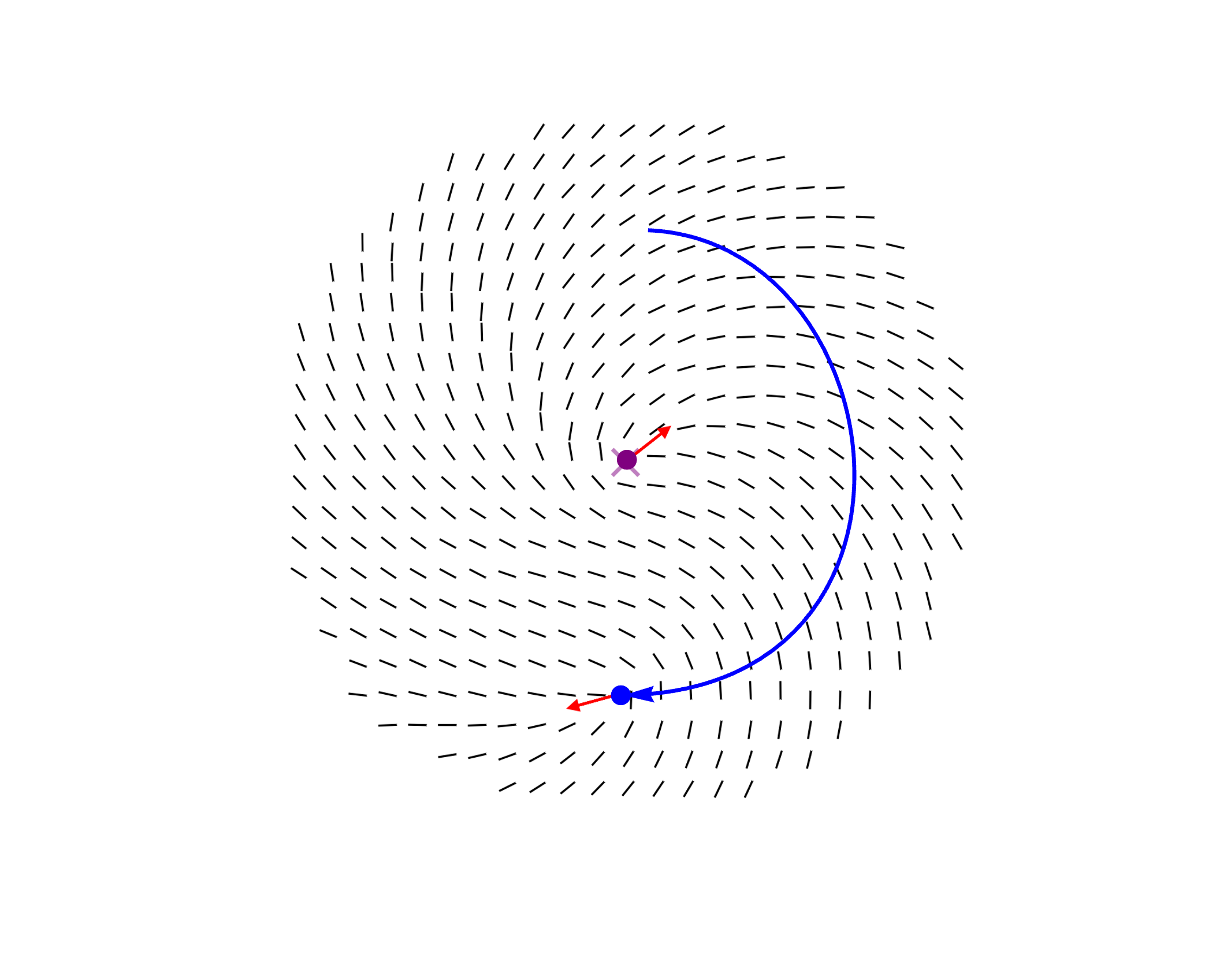}}\\
    \subfloat[]{\includegraphics[width=.25\linewidth]{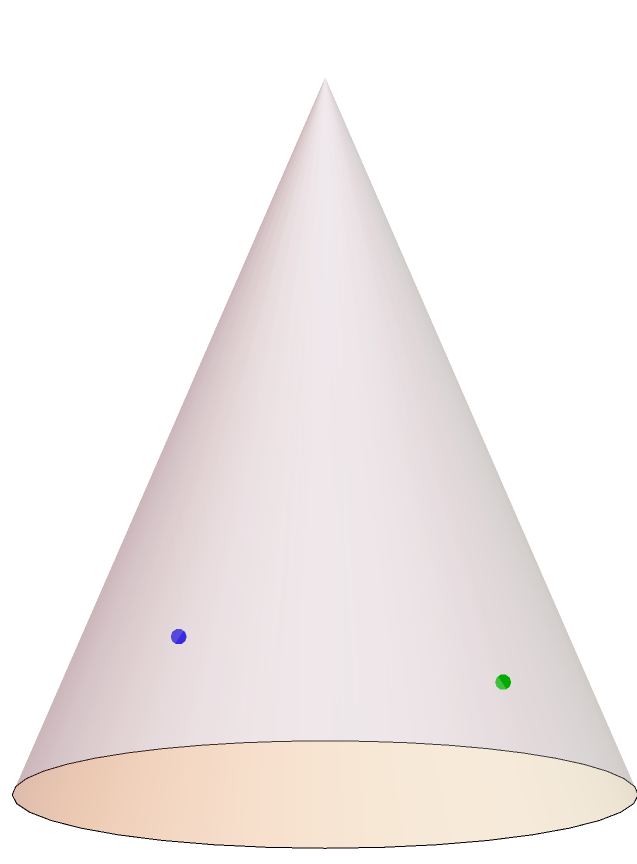}}
    \hfill
    \subfloat[]{\includegraphics[width=.25\linewidth]{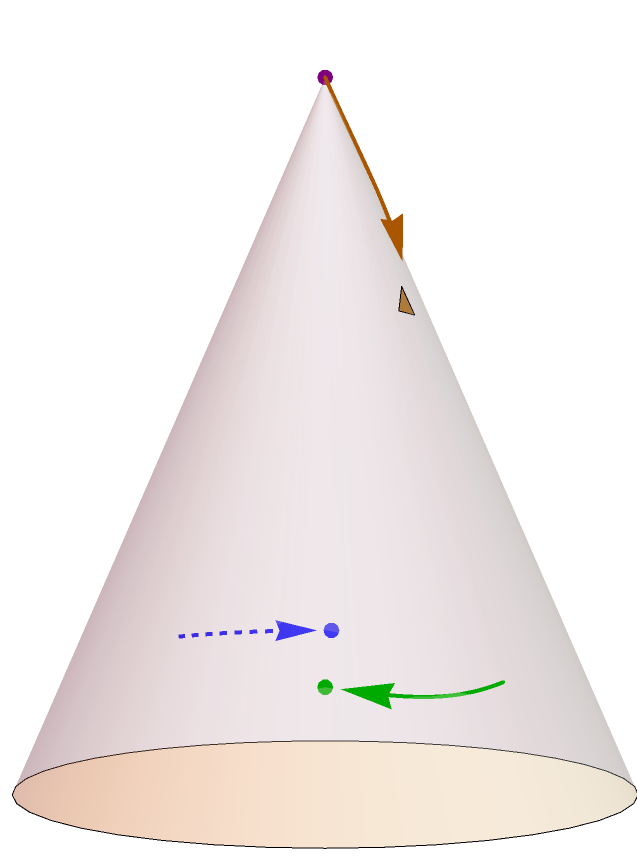}}
    \hfill
    \subfloat[]{\includegraphics[width=.25\linewidth]{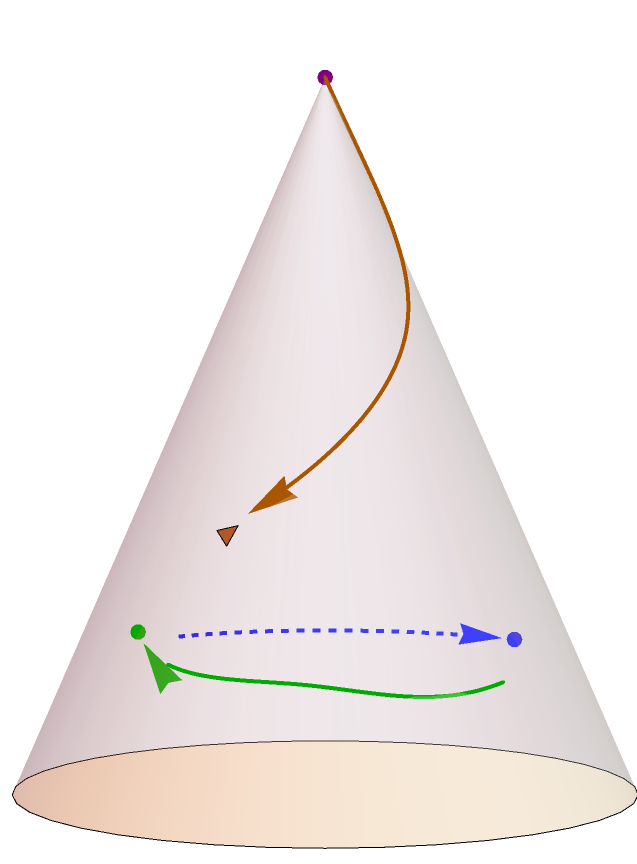}}
    \hfill
    \subfloat[]{\includegraphics[width=.25\linewidth]{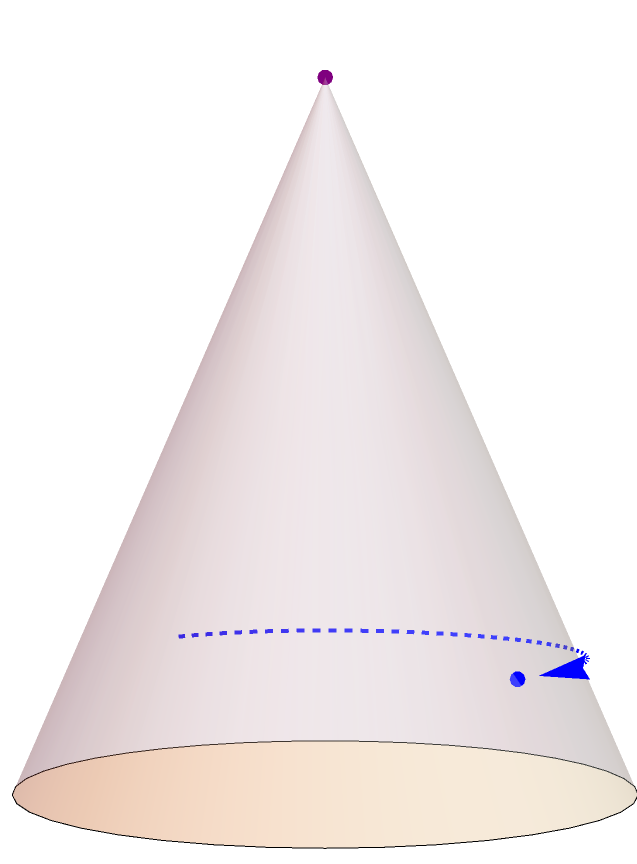}}
    \caption{Key snapshots of simulations depicting the mechanism of defect nucleation at the apex followed by a $-1/2$ defect targeting and annihilating an orbiting $+1/2$ defect. Top row: plots are in isothermal coordinates on a cone with $\chi=0.6$ and $\zeta/(J\gamma^{-1}) = 1.25$ and $\psi=\pi/2$. (a) Initially two $+1/2$ defects are orbiting. (b) a neutral defect pair is nucleated at the apex. (c) the $-1/2$ defect is emitted from the apex and targets one of the orbiting $+1/2$ defects (d) the $-1/2$ defect has successfully annihilated one of the orbiting $+1/2$ defects, and the other orbiting $+1/2$ defect continues to orbit. In all of the plots, the colored dots denote the $+1/2$ defects, the orange triangle denotes the $-1/2$ defect, defects follow the corresponding colored trajectories, the red arrows the polarizations of the $+1/2$ defects, and the purple cross at the origin denotes the apex. In the bottom row are the corresponding plots on a 3D cone.}
    \label{fig:simulation_nucleation}
\end{figure*}

Increasing both the activity and the cone deficit angle reveals an interesting mechanism for $2 \to 1$ defect transition. For sufficiently large $\chi$ and activity, a $\pm 1/2$ defect pair is nucleated at the apex, and then the $-1/2$ defect leaves the apex and annihilates one of the two $+1/2$ flank defects, leaving behind one $+1/2$ defect at the apex and the other $+1/2$ defect on the flanks. See Fig.~\ref{fig:phaseDiagramData} for phase diagram from simulations, and Fig.~\ref{fig:simulation_nucleation} depicting key snapshots of this mechanism for $\psi=\pi/2$.

We do not know analytically when a neutral defect pair must nucleate. However, we can obtain a lower bound on the critical $\chi$ as follows. Suppose a $\pm 1/2$ defect pair was nucleated at the apex. Then from Eq.~\eqref{eq:Forces2DefectsCone} the net force $F$ on the $\sigma = -1/2$ defect is 
\begin{align}
    F &= 2\pi J \left[\chi \left(\sigma - \frac{1}{2}\sigma^2\right) - \sigma^2\right] \nonumber \\
    &= 2\pi J \left[\chi \left(-\frac{1}{2} - \frac{1}{2}\left(-\frac{1}{2}\right)^2\right) - \frac{1}{4}\right] .
\end{align}
The critical $\chi$ is obtained by setting $F=0$, leading to $\chi_c = 2/5$. Thus this mechanism can occur only for $\chi > 2/5$, but it does not have to occur; from our simulations, $\chi_c$ depends on the activity.	

\subsection{Stable cyclic apex-mediated defect pair nucleation, emission, and unbinding}
\label{subsec:cyclic}

\begin{figure*}[t]
    \centering
    \subfloat[]{\includegraphics[width=.16\linewidth, trim={3cm 1cm 3cm 1cm},clip]{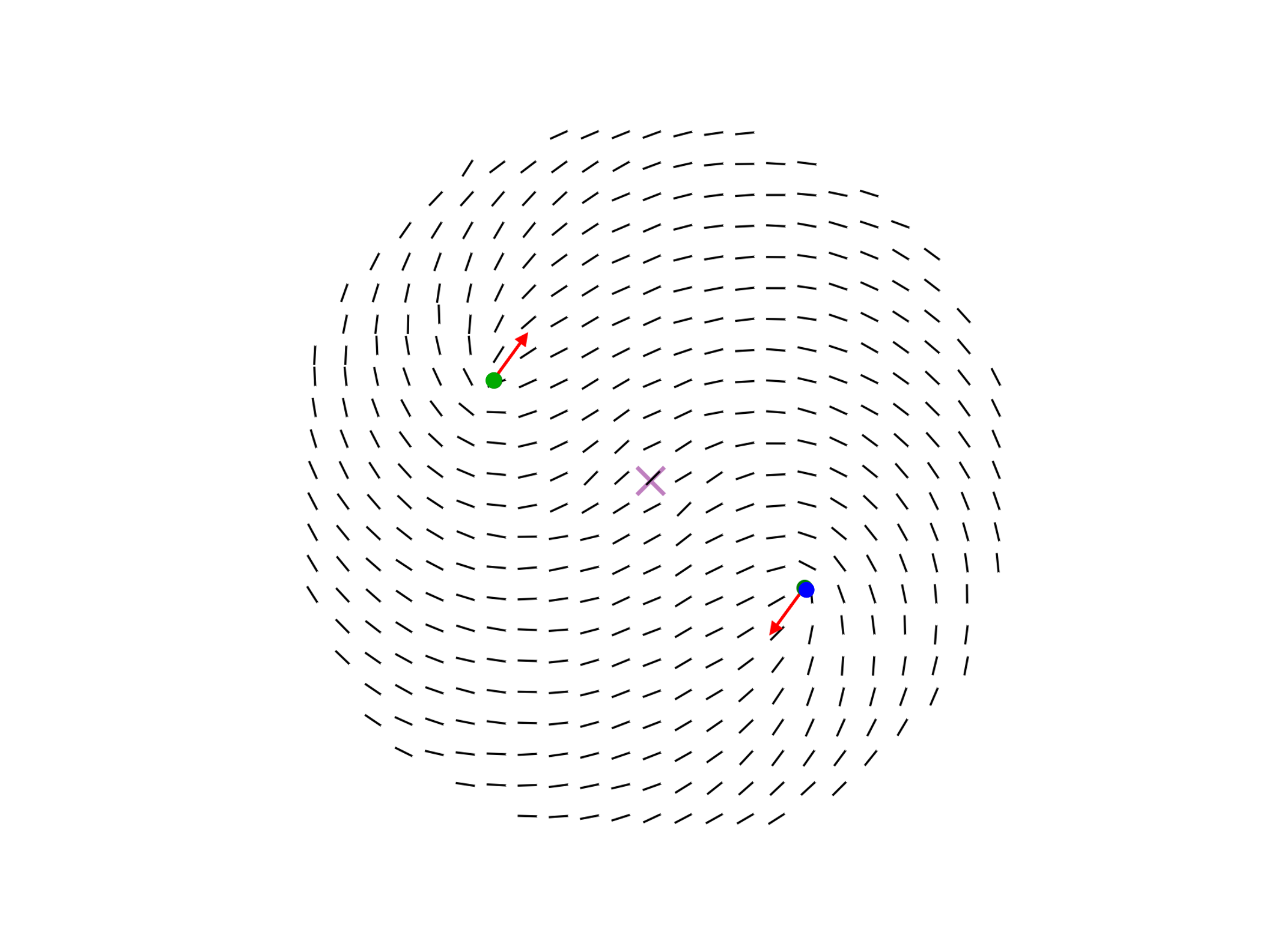}}
    \hfill
    \subfloat[]{\includegraphics[width=.16\linewidth, trim={3cm 1cm 3cm 1cm},clip]{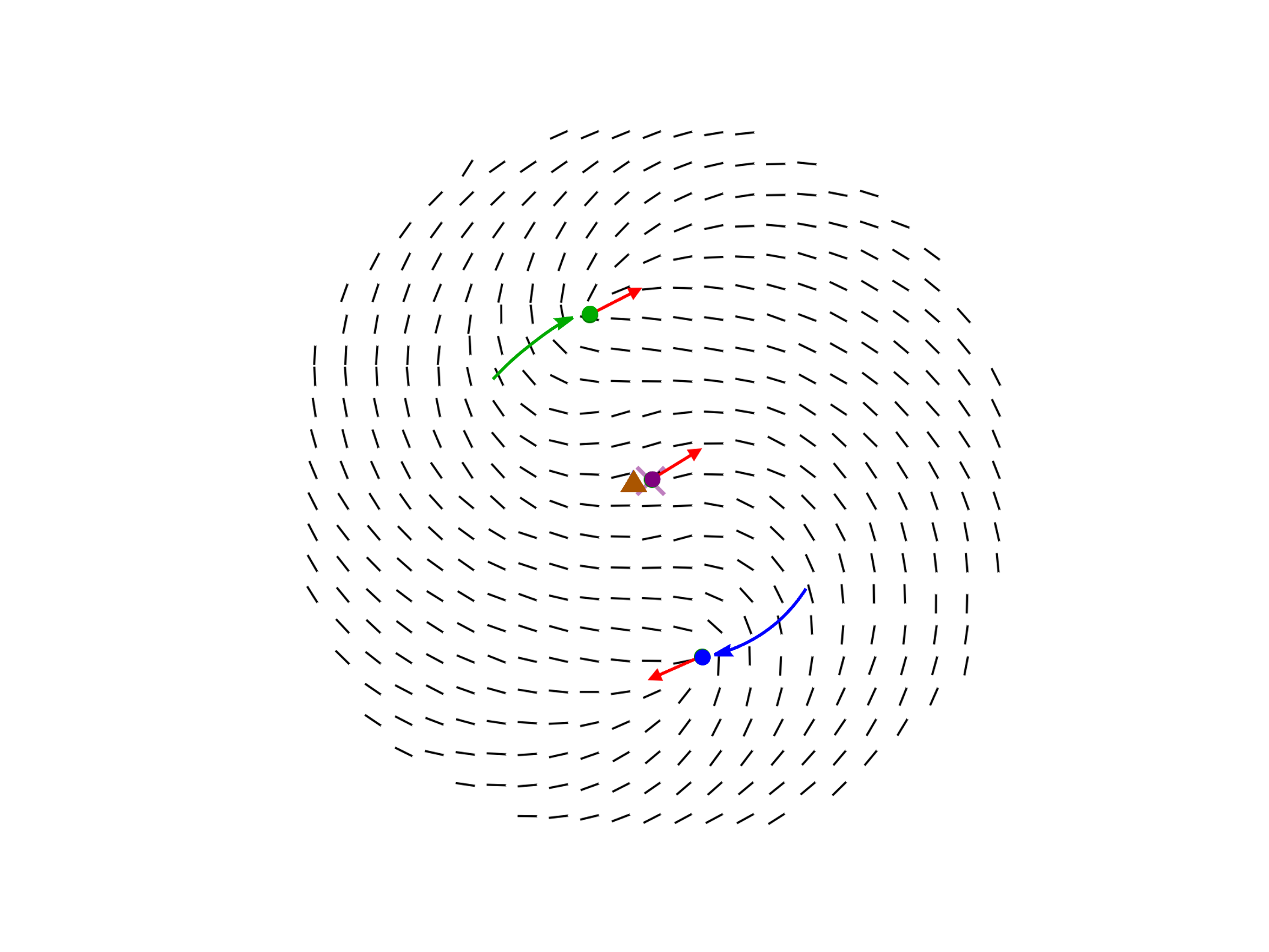}}
    \hfill
    \subfloat[]{\includegraphics[width=.16\linewidth, trim={3cm 1cm 3cm 1cm},clip]{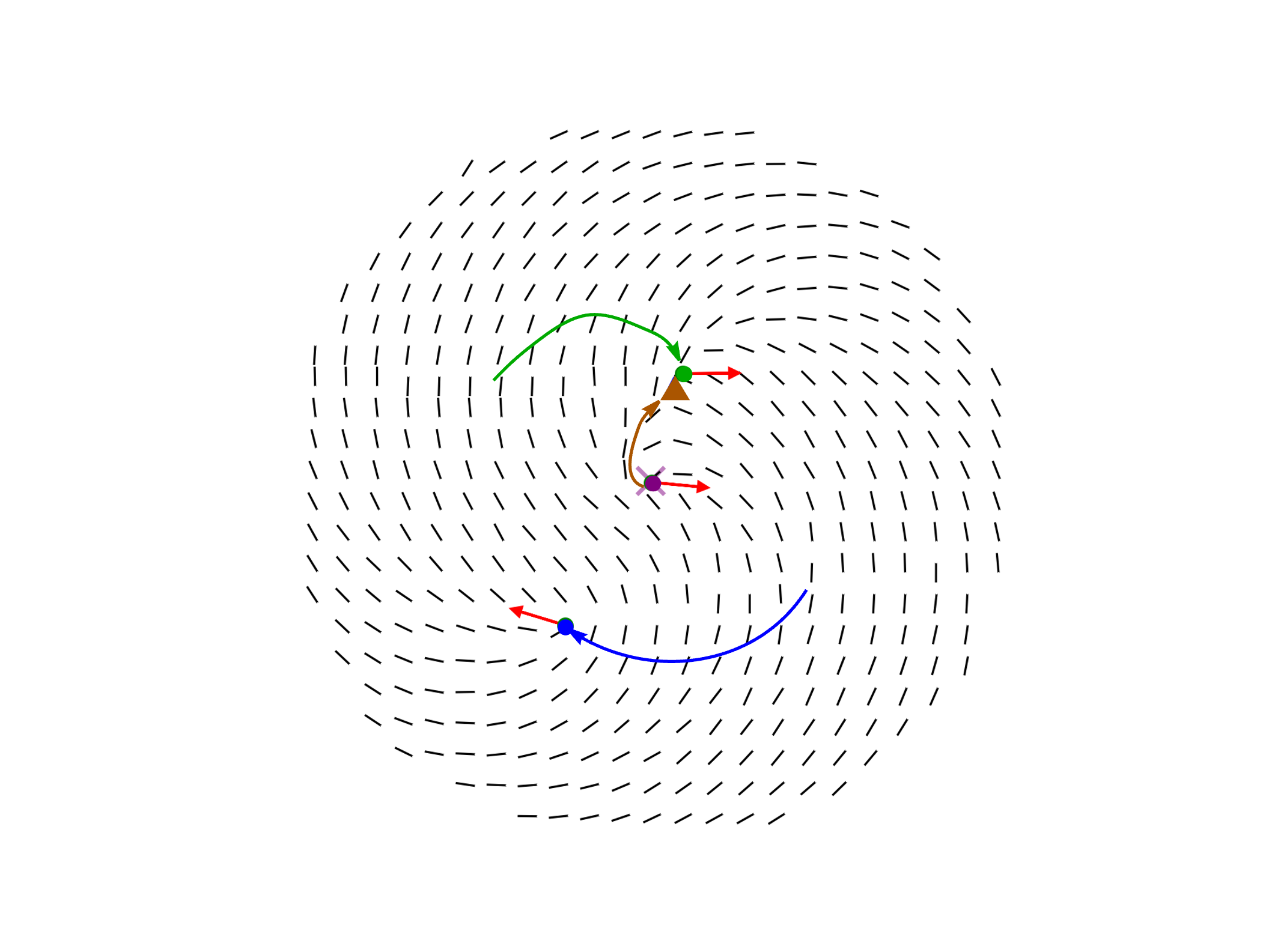}}
    \hfill
    \subfloat[]{\includegraphics[width=.16\linewidth, trim={3cm 1cm 3cm 1cm},clip]{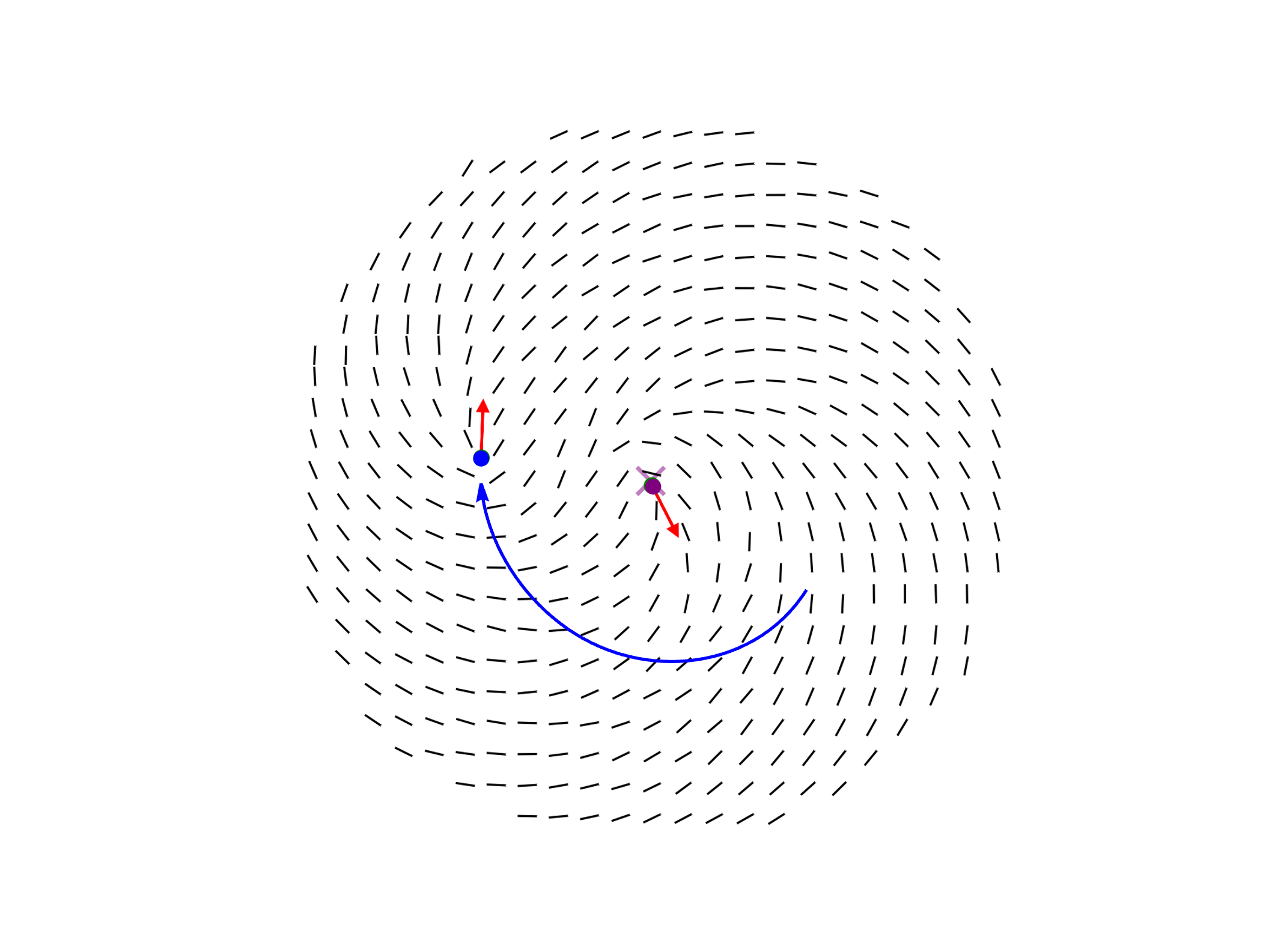}}
    \hfill
    \subfloat[]{\includegraphics[width=.16\linewidth, trim={3cm 1cm 3cm 1cm},clip]{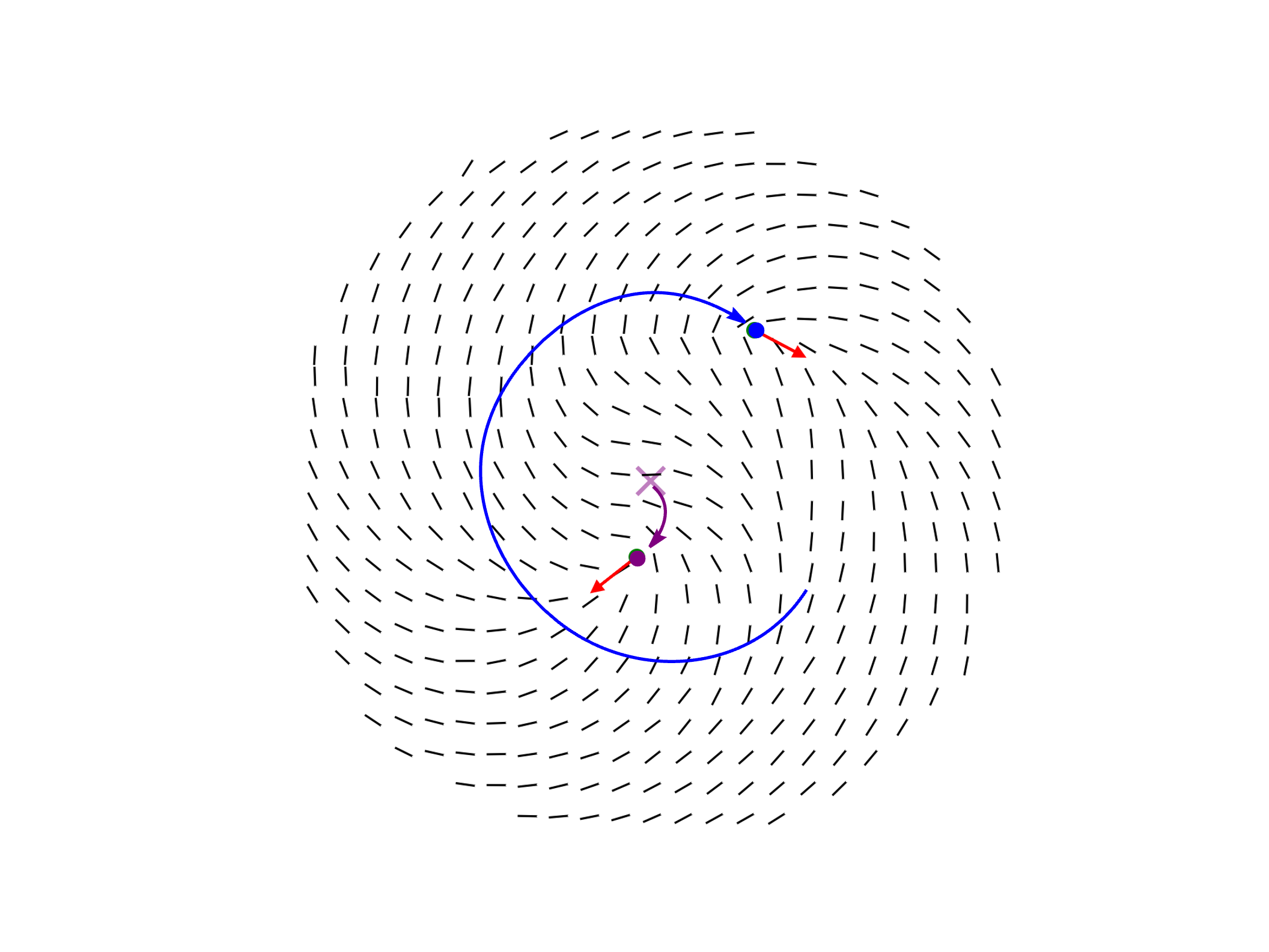}}
    \hfill
    \subfloat[]{\includegraphics[width=.16\linewidth, trim={3cm 1cm 3cm 1cm},clip]{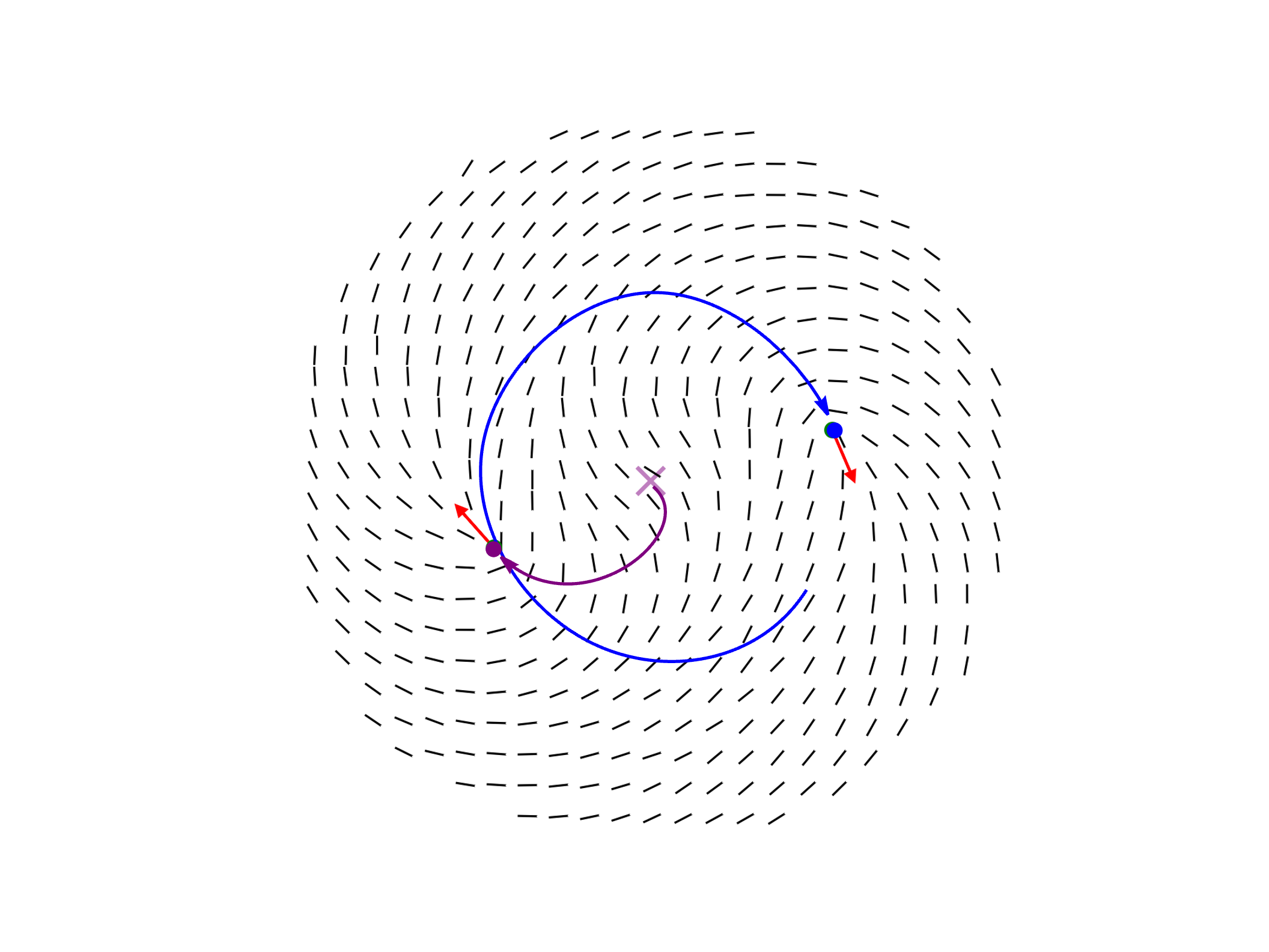}}\\
    \subfloat[]{\includegraphics[width=.16\linewidth]{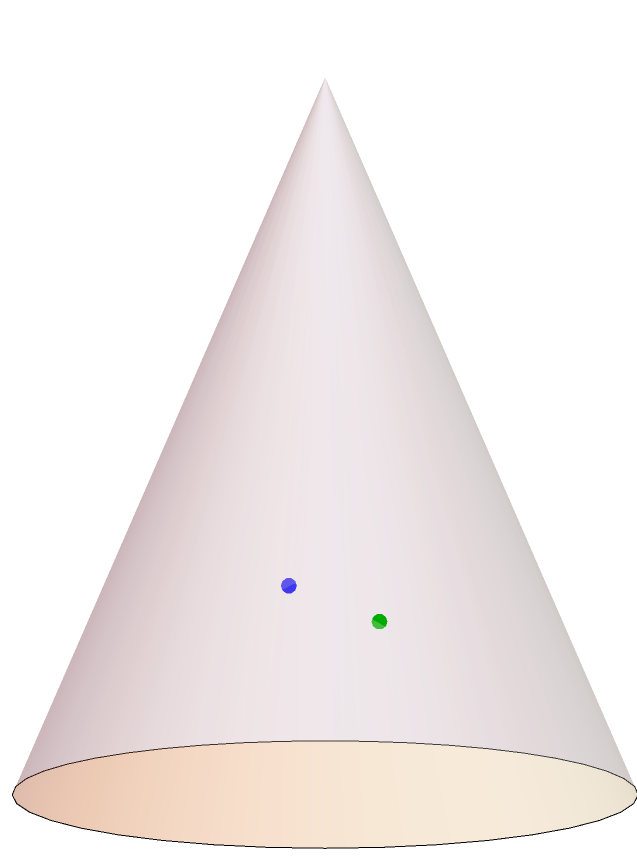}}
    \hfill
    \subfloat[]{\includegraphics[width=.16\linewidth]{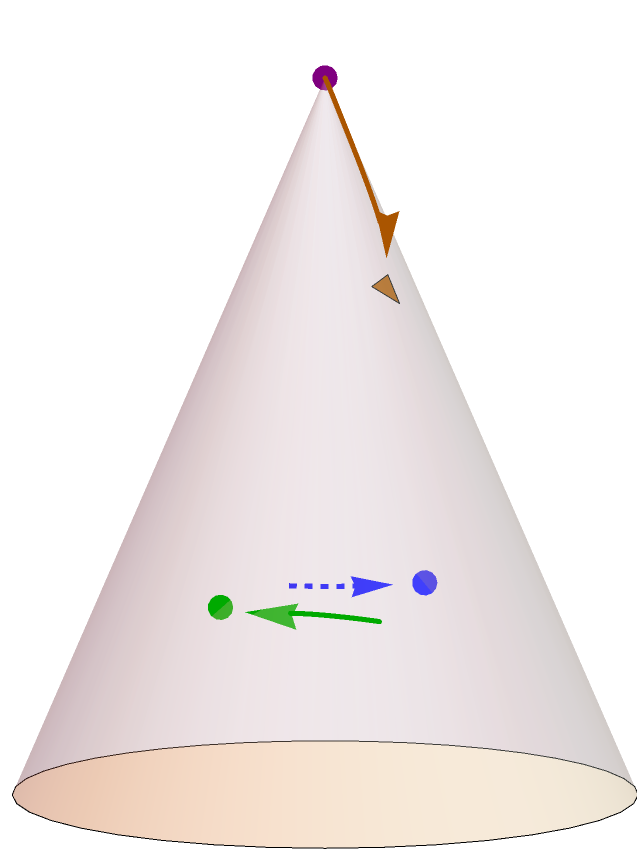}}
    \hfill
    \subfloat[]{\includegraphics[width=.16\linewidth]{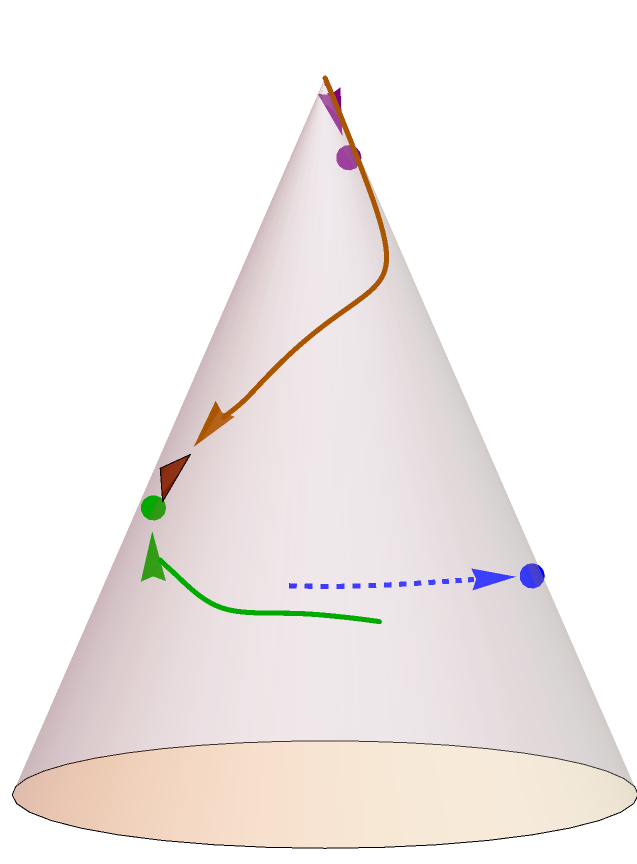}}
    \hfill
    \subfloat[]{\includegraphics[width=.16\linewidth]{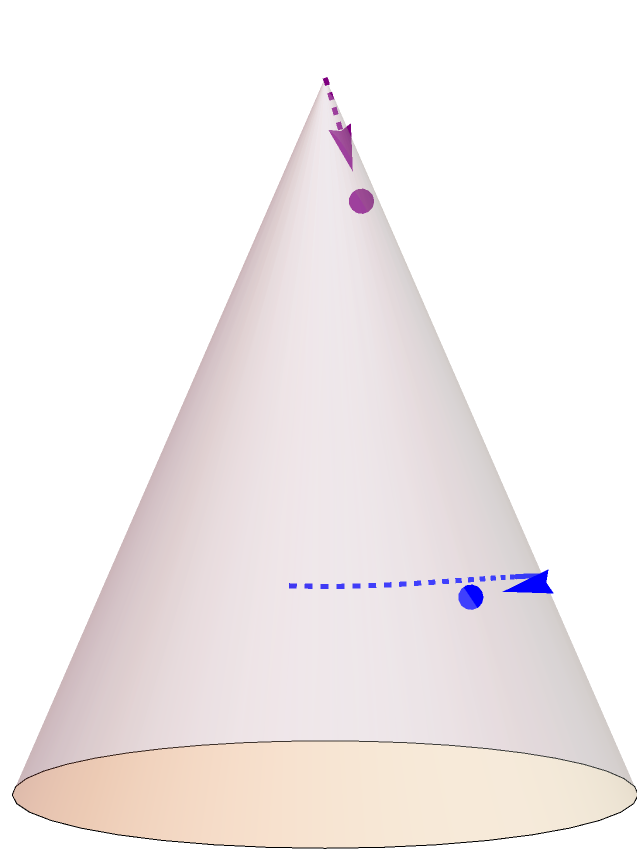}}
    \hfill
    \subfloat[]{\includegraphics[width=.16\linewidth]{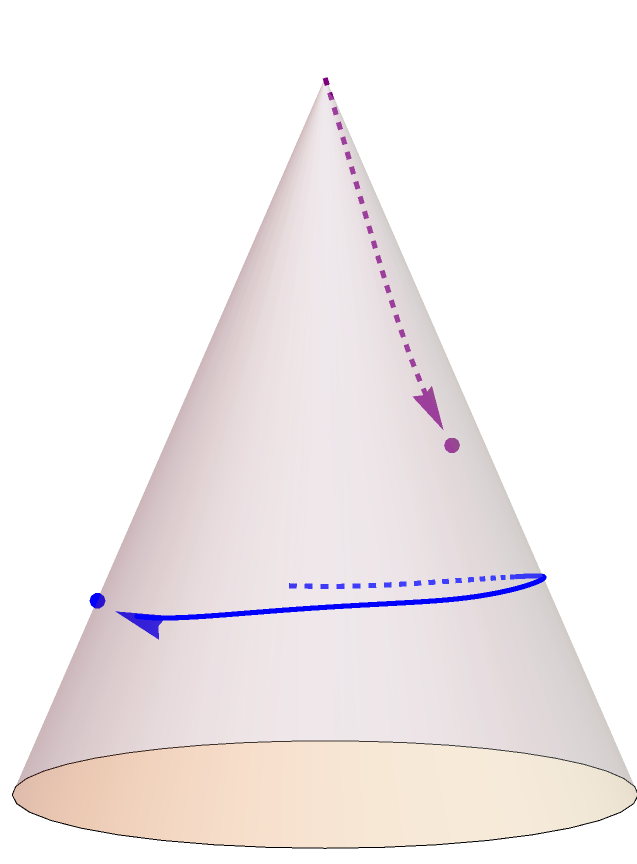}}
    \hfill
    \subfloat[]{\includegraphics[width=.16\linewidth]{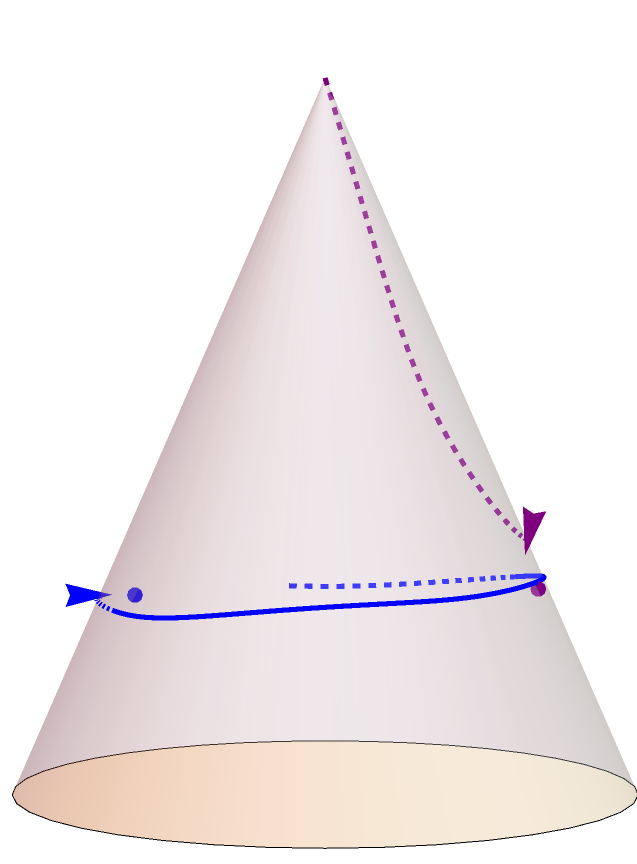}}
    \caption{Key snapshots of simulations depicting cyclic $2\to 1\to 2 \ldots$ defect transition mechanism. Top row: plots are in isothermal coordinates on a cone with $\chi=0.6$ and $\zeta/(J\gamma^{-1}) = 2$ and $\psi=\pi/2$. (a) Initially two $+1/2$ defects are orbiting. (b) a neutral defect pair is nucleated at the apex. (c) the $-1/2$ defect is emitted from the apex and targets one of the orbiting $+1/2$ defects (d) the $-1/2$ defect has successfully annihilated one of the orbiting $+1/2$ defects, and the other orbiting $+1/2$ defect continues to orbit. (e) the $+1/2$ defect begins to unbind from the apex (f) the $+1/2$ defect has successfully unbound (escaped) from the apex. In all of the plots, the colored dots denote the $+1/2$ defects, the orange triangle the $-1/2$ defect,  defects follow the corresponding colored trajectories, the red arrows the polarizations of the $+1/2$ defects, and the purple cross at the origin denotes the apex. In the bottom row are the corresponding plots on a 3D cone.}
    \label{fig:simulation_cyclic}
\end{figure*}

Finally, simulations show that further increase in activity and deficit angle results in exotic cyclic defect pair nucleation, emission, and unbinding (see Fig.~\ref{fig:simulation_cyclic} for key snapshots depicting this exotic behavior for $\psi=\pi/2$), which can be qualitatively explained as follows.
Previously, we have seen that for sufficiently large deficit angle $2\pi\chi$, defect nucleation and emission can occur, causing a transition from a two defect orbit to a stable one defect orbit, i.e. a $2 \to 1$ defect transition. On the other hand, we have also seen that for sufficiently large activity $\zeta$, a single defect orbit can become a stable circular two defect orbit by defect unbinding from the apex, i.e. a $1\to 2$ defect transition. Thus it is natural to expect for sufficiently large $\chi$ and $\zeta$, we can have both of these transitions happen successively. The full phase diagram in activity-cone deficit angle based on the simulation results is provided in Fig.~\ref{fig:phaseDiagramData} summarizing the main dynamical states and transitions between them.

\section{Discussion}
\label{sec:discussion}

In this work, we have investigated the dynamics of a compressible, overdamped active nematic on disks and cones, with a focus on the role of boundary conditions. By imposing strong anchoring boundary conditions at the base of a disk or a cone, we have uncovered a rich phase diagram of circular orbits of one or two $+1/2$ flank defects, with transitions between these states mediated by defect absorption, defect unbinding, and defect pair nucleation at the apex. Strong anchoring boundary conditions with the director at a $45^\circ$ angel to the cone base are particularly interesting. Moreover, the Born-Oppenheimer approximation (nematic textures instantaneously follow defect positions) allows us to make analytical predictions about the dynamics of active nematic defects on curved geometries. Many of these predictions were indeed corroborated by our full numerical simulations. For example, balancing the Coulomb forces (including those from image charges) with the motile active force led to stable circular orbits in the regime of low activity and deficit angle, at locations predicted by the theory. And at higher activity, the prediction for the critical activity for defect unbinding from the apex agreed with theory. Going beyond what can be derived from a simple Born-Oppenheimer approximation, the full numerical simulations revealed that curvature, acting as a lightning rod with activity serving as the catalyst, can induce the nucleation of neutral defect pairs, with the emission of mobile $-1/2$ defects, despite their local 3-fold symmetry. This finding provides a route and possible explanation for defect pair nucleation in more general curved active systems.

We now comment on the relation of our work to other interesting investigations~\cite{norton2018insensitivity,opathalage2019self,joshi2023disks}, which considered the related model of an incompressible nematic on a disk or an annulus. Although there is no geometric charge at the center, as occurs at the apex for cones with $\chi > 0$, Ref.~\cite{norton2018insensitivity} via simulations found that for sufficiently large activity, a two defect orbit with tangential boundary conditions can become stable via defect pair nucleation at the origin followed by defect emission of the $-1/2$ defect which then annihilates one of the original orbiting $+1/2$ defects, similar to the mechanism discussed in Sec.~\ref{subsec:nucleation}. Experiments on a disk in Ref.~\cite{opathalage2019self} show that $2\to 1\to 2 \ldots$ transitions occur via defect pair nucleation at the boundary followed by annihilation and then again defect pair nucleation, etc, similar to our mechanism in Sec.~\ref{subsec:cyclic}, with the intriguing difference that defect pair nucleation occurs at the boundary vs the apex for our case. Finally, Ref.~\cite{joshi2023disks} numerically found a phase diagram in the case of a disk and annulus, which when restricted to the regime where the inner radius of the annulus is small compared to the outer radius, agrees with our phase diagram (Fig.~\ref{fig:phaseDiagramData}) for small deficit angle: for small activity, circular orbits are allowed, and for large activity, the dynamics is chaotic. The results of this specific annulus geometry is not surprising since it is akin to a truncated cone, with free boundary conditions imposed at the truncation~\cite{vafa2022defectAbsorption}. An inverted truncated cone is like a banked racetrack, and it would be interesting to study active nematic dynamics on such a structure as a function of apex ratio, with an unquantized Gaussian curvature residing inside the truncation.

In the future, it would be worth studying the dynamics of active matter coating a \emph{hyperbolic} cone. Preliminary investigations of ground states have revealed that even in the passive setting, extra neutral defect pairs leading to additional $+1/2$ flank defects are nucleated at the apex, depending on the deficit angle~\cite{vafa2023}. It would also be worth exploring further the regime of large activity, where we expect active turbulence, interacting with a delta function of Gaussian curvature, leading to an active analog of the Debye-H\"{u}ckel screening problem of electrolytes in two dimensions.

Finally, it is worth noting the possible experimental setups that could test our theoretical and computational predictions. It has already been demonstrated that by using a rotating cuvette containing a yield-stress fluid, one can form toroidal droplets on which mixtures of microtubule-kinesin motor mixtures are stabilized and nucleate active nematic topological defects~\cite{ellis2018curvature}. Extending this elegant experimental technique to stabilize active nematic microtubule-motor protein mixtures on conical droplets might be challenging, but could be an ideal test case for exploring active defect dynamics in a controlled manner. By tuning the sharpness of the cone in yield-stress droplets of various shapes, and the activity through varying ATP concentration, one could probe the phase diagram presented in Fig.~\ref{fig:phaseDiagram}. Similarly, the possibility of forming cell monolayers on corrugated surfaces has been demonstrated using polyacrylamide hydrogels fabrication by UV photocrosslinking~\cite{luciano2021cell}. This technique can be easily adapted to achieve cell monolayers on conical surfaces of varying sharpness. We expect especially complex dynamics when $\chi$ moves closer to 1 and the activity becomes large. One could then probe dynamics of elongated, weakly adhesive mesenchymal cells such as mouse fibroblast to explore the rich phase diagram studied here. Indeed, such cell layers were already instrumental to probe defect locations in confined disks~\cite{duclos2017topological}. Confining defects to the cone, together with our theoretical and computational predictions, could result in one of the first quantitative assessments of topological defect unbinding in active matter with a controlled environment. In addition to subcellular filaments and cell layers, important examples of living liquid crystals, embedding active bacteria in pre-patterned liquid crystal textures~\cite{turiv2020polar} could be a fertile ground for testing active nematic defects on cones as examples of non-trivial curvature singularities. Indeed, such living liquid crystals have been shown to exhibit active nematic defects~\cite{zhou2014living}, and the possibility of forming passive liquid crystals on conical geometries has been demonstrated~\cite{lagerwall2023liquid}. As such, living liquid crystals on a cone could provide another controlled means for obtaining dynamically ordered defect structures and studying details of defect emission, nucleation, and unbinding in active matter, as explored in this paper.

\acknowledgments{This work is partially supported by the Center for Mathematical Sciences and Applications at Harvard University (F. V.), and by the Harvard Materials Research Science and Engineering Center via Grant DMR-2011754 (D.R.N.). A. D. acknowledges funding from the Novo Nordisk Foundation (grant No. NNF18SA0035142 and NERD grant No. NNF21OC0068687), Villum Fonden (Grant no. 29476), and the European Union (ERC, PhysCoMeT, 101041418). Views and opinions expressed are however those of the authors only and do not necessarily reflect those of the European Union or the European Research Council. Neither the European Union nor the granting authority can be held responsible for them.}

\bibliography{refs}
\end{document}